\documentclass[manuscript]{aastex}
\usepackage{epstopdf}
\usepackage{subfigure}
\usepackage{amsmath}

\usepackage{longtable}

\shorttitle{GOES class}
\shortauthors{Reep, Bradshaw, McAteer}

\begin{document}

\title{On the sensitivity of the GOES flare classification to properties of the electron beam in the thick target model}

\author{J. W. Reep}
\affil{Department of Physics and Astronomy, Rice University, Houston, TX 77005, USA}
\email{jeffrey.reep@rice.edu}\and
\author{S. J. Bradshaw}
\affil{Department of Physics and Astronomy, Rice University, Houston, TX 77005, USA}
\email{stephen.bradshaw@rice.edu}\and
\author{R.T.J. McAteer}
\affil{Department of Astronomy, New Mexico State University, Las Cruces, NM 88003 USA}
\email{mcateer@nmsu.edu}

\begin{abstract}

The collisional thick target model, wherein a large number of electrons are accelerated down a flaring loop, can be used to explain many observed properties of solar flares.  In this study, we focus on the sensitivity of GOES flare classification to the properties of the thick target model.  Using a hydrodynamic model with RHESSI-derived electron beam parameters, we explore the effects of the beam energy flux (or total non-thermal energy), the cut-off energy, and the spectral index of the electron distribution on the soft X-rays (SXRs) observed by GOES.  We conclude that (1) the GOES class is proportional to the non-thermal energy $E^{\alpha}$ for $\alpha \approx 1.7$ in the low energy passband (1-8 \AA) and $\approx 1.6$ in the high energy passband (0.5-4 \AA); (2) the GOES class is only weakly dependent on the spectral index in both passbands; (3) increases in the cut-off will increase the flux in the 0.5-4 \AA\ passband, but decrease the flux in the 1-8 \AA\ passband, while decreases in the cut-off will cause a decrease in the 0.5-4 \AA\ passband and a slight increase in the 1-8 \AA\ passband.

\end{abstract}


\section{Introduction}

One of the outstanding problems in solar physics is to identify the physical mechanism that gives rise to solar flares.  The electron beam heating scenario has been a popular mechanism to explain X-ray emissions in solar flares.  Under this mechanism, electrons accelerated high in the corona stream down a coronal loop towards the chromosphere, depositing their energy through collisions, and emitting high-energy X-rays via electron-ion bremsstrahlung with the ambient ions \citep{brown1971,emslie1978}.  As they deposit their energy, the footpoints heat up and chromospheric ablation drives material into the corona, filling the loop and producing lower energy X-rays via thermal bremsstrahlung \citep{fisher1985a,fisher1985b,fisher1985c}.  

Solar flares are commonly classified in terms of their peak SXR flux in the 1-8 \AA\ band, as observed by the Geostationary Operational Environmental Satellite (GOES) X-ray flux monitor at terrestrial distance \citep{garcia1994}.  The largest flares, X-class, have peak fluxes greater than 10$^{-4}$ W m$^{-2}$, with smaller flares classified by a decrease in peak flux by factors of 10 as M, C, B, and A-class.  The flux in this GOES passband is determined primarily by a combination of thermal and non-thermal bremsstrahlung, and is therefore connected intimately with the properties of the beam of accelerated electrons.  

A large number of flares (but not all) show the Neupert effect, which states that the fluence of the hard X-rays (HXRs) is proportional to the flux of the SXRs, or equivalently, that the HXR flux is proportional to the time derivative of the SXR flux \citep{neupert1968,dennis1993,veronig2005}.  In the thick-target model, the deposition of energy by the electron beam induces the observed HXR bursts, subsequently ablating material into the corona and heating the plasma to produce the more gradual SXR light curves.  \citet{lee1995} demonstrate that the Neupert effect, combined with this model of chromospheric ablation, should lead to a linear proportionality between the non-thermal energy and the SXR flux; in contradiction to this, \citet{warren2004} show that the SXR flux (as seen by GOES) should be proportional to non-thermal energy $E^{1.75}$.  

In the last decade the Reuven Ramaty High-Energy Solar Spectroscopic Imager (RHESSI; \citealt{lin2002}) and more recently the Fermi Gamma-Ray Burst Monitor (GBM; \citealt{meegan2009}) have provided excellent coverage of soft and hard X-ray spectra in flares.  The spectra obtained with these instruments can be used to derive the properties of electron beams through various inversion methods \citep{brown2003,kontar2011}.  The basic method involves inverting the equation relating an observed intensity to parameters of the electron beam (see Equation \ref{thicktargetbrem}, below) to solve for the mean electron distribution function (see for example, \citealt{holman2003, kontar2003, piana2003}).  \citet{brown2006} evaluate four different methods of inversion, concluding that all of the methods recover the general magnitude of a given distribution but have trouble recovering sharp features.

Electron beams are primarily characterized by three parameters: the energy flux of the electrons, the spectral index of the electron distribution, and the low-energy cut-off of that distribution.  The primary hindrance in determining the electron distribution accurately is the low-energy cut-off.  At low energies, thermal bremsstrahlung masks the non-thermal component, rendering accurate determination of the cut-off difficult \citep{holman2011,kontar2011}.  Further, most studies assume a simple power-law with a sharp cut-off for the electron distribution, although it is not clear that this model is correct \citep{saint2005,holman2011}.  

Numerous models have been developed to study heating processes in solar flares.  \citet{nagai1980} developed a 1D hydrodynamic model of loops heated by a thermal conduction front to study the formation of SXR emission in flares.  Later models, using the heating function derived by \citet{emslie1978}, explored the hydrodynamic response of the solar atmosphere to a non-thermal beam of electrons (\citealt{nagai1984,macneice1984,mariska1989}).  Most of these studies assumed heating lasted for less than a minute, with fixed beam parameters.  However, using results from RHESSI, we can now develop a model combining time-dependent beam properties with an advanced hydrodynamic model to forward model spectra and study the effects of an electron beam on observed spectra directly.  

In this paper, we examine the sensitivity of the GOES classification to the beam properties.  We use a numerical model to explore the effect of varying each of the beam parameters on light curves as measured by GOES.  In Section \ref{hydrad} we briefly describe the numerical code used to perform the simulations and several improvements that make it suitable for the studies undertaken here.  In Section \ref{bhf} we describe the beam heating function and the specific assumptions we have made in implementing it, and in Section \ref{brems} we describe the calculation of bremsstrahlung emissions and the synthesis of GOES light curves from the simulations.  In Section \ref{results} we present the results of 60 numerical experiments to explore the correlation of GOES classification to each beam parameter derived for two flares observed by RHESSI.  Finally, we summarize the results and discuss future work in Section \ref{conclusions}.

\section{Numerical Modeling}

\subsection{The HYDRAD code}
\label{hydrad}

The work presented here has been performed using the HYDRAD code, which solves the hydrodynamic equations for an isolated magnetic flux tube and a multi-fluid plasma \citep{bradshaw2013}.  The equations (conservation of mass, momentum, and energy) and assumptions are detailed in the appendix of \citet{bradshaw2013}.  However, there have been a number of important improvements to the code.

\begin{figure}
\centering
\includegraphics[width=4.25in]{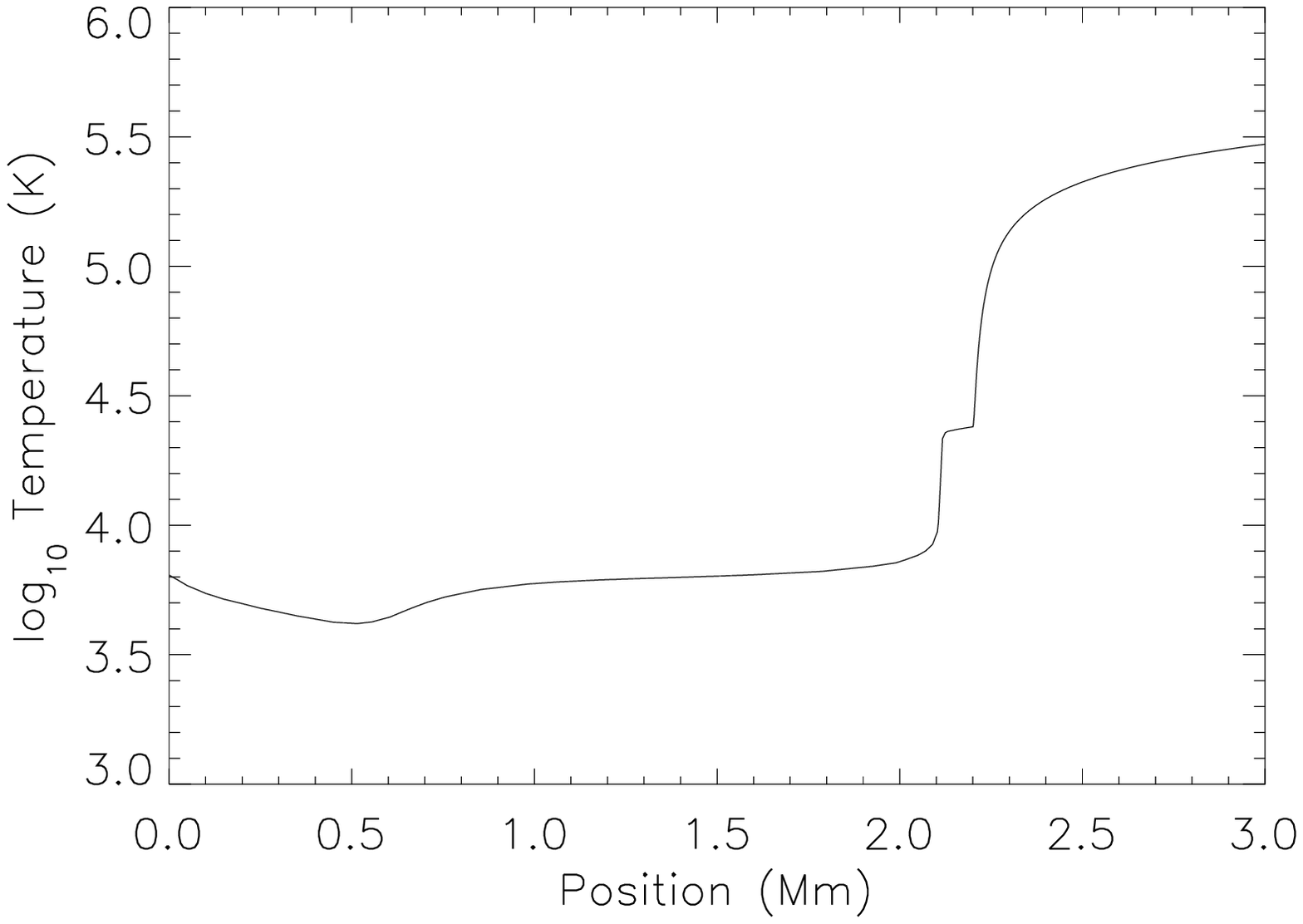}
\hspace{0.1in}
\centering
\includegraphics[width=4.25in]{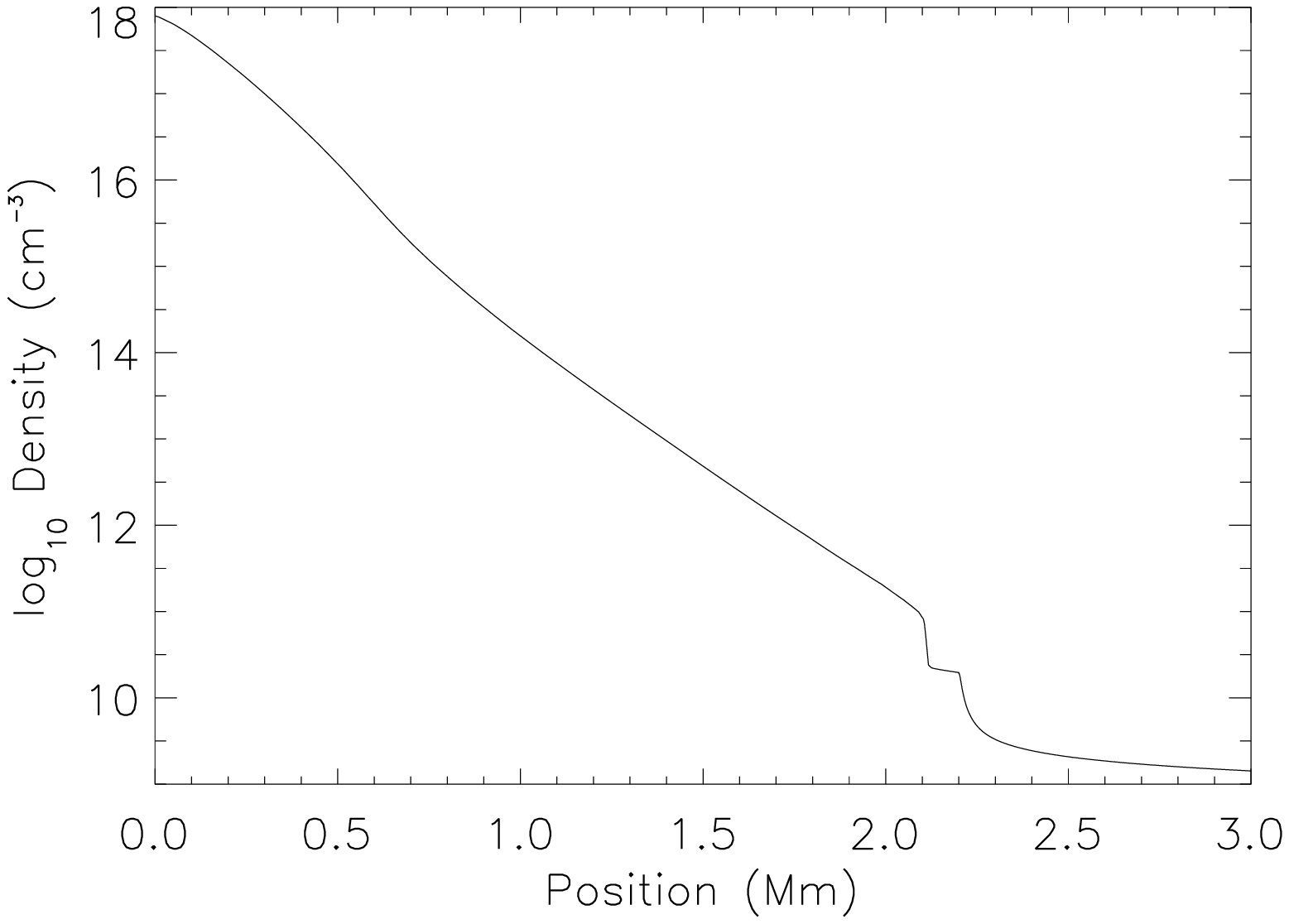}
\caption{{\it Top:} The temperature profile in the lower atmosphere at the start of Run 1 (in the field-aligned direction).  The hydrogen and electron temperatures were initially equal.  {\it Bottom:} The density profile in the lower atmosphere at the start of Run 1.  \label{loweratm}}
\end{figure}

First, the pre-flare atmosphere is now based on the VAL model C of the photosphere and chromosphere \citep{vernazza1981}, allowing for more realistic temperature and density profiles in the lower solar atmosphere.  We recalculated the density distribution for hydrostatic equilibrium to be consistent with the average particle mass chosen for our model ($m_i = 2.171 \times 10^{-24}$~g), which accounts for the relative abundances of hydrogen, helium, and heavier elements (roughly 90\% hydrogen and 10\% helium).  The initial transition region and corona are derived by integrating the hydrostatic equations from the top of the VAL C atmosphere to the apex of the coronal loop.  Second, the code has been modified to allow for the presence of neutrals in the lower atmosphere.  The two fluids in the code are now electrons and hydrogen atoms (which includes ions and neutrals), with trace amounts of electrons due to ionization of heavier elements.  The energy equations have been modified in a manner similar to \citet{macneice1984} to include the ionization of hydrogen to account for the effects of neutrals on energy balance between the two fluids (e.g., via collisions and thermal conduction of neutrals, \citealt{orrall1961}).  Finally, the chromospheric radiative energy balance is now based on the recipe derived by \citet{carlsson2012}, which accounts for cooling from optically thick lines and continua, heating due to the same processes, and heating due to coronal radiation (back-warming from the hot corona).  We also used their ionization balance to calculate the electron number density ($n_e$), for consistency with this radiation calculation.  Figure \ref{loweratm} shows the lower atmosphere density and temperature profiles at the start of Run 1.

The loop geometry is semi-circular, along the field-aligned direction.  The initial temperature and density profiles were found by integrating the hydrostatic equations from the chromosphere (VAL model C, \citealt{vernazza1981}) to the apex of the coronal loop (as done, for example, in \citealt{aschwanden2001}).  The electron and ion populations were assumed to be in thermal equilibrium at the beginning of simulations.  The loops are cool and tenuous, before significant heating drives chromospheric ablation (compare, {\it e.g.}, \citealt{macneice1984}, \citealt{nagai1984}, \citealt{mariska1989}).

\subsection{Beam Heating Function}
\label{bhf}

We have implemented a modified form of the beam heating function derived by \citet{emslie1978}.  Following \citet{mariska1989} and \citet{li1991}, we add a low-energy knee to the beam electron distribution while keeping the total beam energy constant (also compare the simulations of \citealt{warren2006}).  As those authors note, this form of the electron distribution produces smoother temperature and density variations than a sharp cut-off, without significantly altering the X-ray spectrum (see Sections 3.3 and 3.4 of \citealt{holman2011} for comparisons of different shapes of low-energy cut-offs).  The distribution function becomes:

\begin{equation}
\mathfrak{F}(E_{0}, t) = \frac{4 (\delta - 2) F_{0}(t)}{(\delta + 2) E_{c}^{2} }
  \begin{cases}
   (\frac{E_{0}}{E_{c}})^{2} & \text{if } E_{0} \leq E_{c} \\
   (\frac{E_{0}}{E_{c}})^{- \delta}       & \text{if } E_{0} \geq E_{c}
  \end{cases}
\label{distfunc}
\end{equation}

\noindent where $\delta$ is the spectral index above the cut-off energy, $E_{c}$, and $F_{0}(t)$ is the energy flux of the beam as a function of time (erg sec$^{-1}$ cm$^{-2}$).  The usual caveats apply to this distribution: the number of electrons accelerated is extremely high, there may be instabilities due to Langmuir wave generation, and there may be a return current generated (see \citealt{brown1977} for more information).  Following \citet{li1991}, the heat input as a function of position and time is:

\begin{equation}
H_{e}(s, t) = \frac{4 \pi e^{4} n_{H} \gamma (\delta - 2)}{\mu_{0} (\delta + 2)} \frac{F_{0}(t)}{E_{c}^{2}} 
	\begin{cases}
		z^{- \delta / 2} B[\frac{\delta}{2}, \frac{2}{4 + \beta}]  & \text{if } z > 1 \\
		z \int_{z}^{1} \frac{ dy }{y^{2} (1 - y)^{2/3}} + z^{- \delta / 2} \int_{0}^{z} y^{\delta / 2 - 1} (1 - y)^{- 2 / 3} dy & \text{if } z \leq 1
	\end{cases}
\end{equation}

\noindent where $n_{H}$ is the hydrogen density, $\mu_{0}$ the cosine of the pitch angle of injection, $\gamma = x \Lambda + (1 - x) \Lambda^{\prime}$, $\beta = [2 x \Lambda + (1 - x) \Lambda^{\prime \prime}] / [\Lambda^{\prime} + x (\Lambda - \Lambda^{\prime})]$ (see appendix \ref{coulomb} for explanation of the different Coulomb logarithms $\Lambda$ used here), $x$ is the ionization fraction of hydrogen, $z = N / N_{c}$ for $N$ the column density and $N_{c} = \mu_{0} E_{c}^{2} / [ (2 + \beta / 2) \gamma 2 \pi e^{4}]$ the stopping depth of an electron with energy $E_{c}$, and finally $B$ is the (complete) beta function.  

Following the ideas of \citet{hawley1994}, we also generalize this expression to non-uniform ionization, which is important for recovering spectral breaks in observed spectra.  As noted by those authors, the heating rate is fairly insensitive to the quantity $\beta$, but it must be constant to integrate Equation 26 of \citet{emslie1978}.  Therefore, we assume that $\beta = 2$ (corresponding to a fully ionized plasma, although we do {\it not} assume $x=1$ in general) and integrate.  Second, we modify the column depth to take into account the varying ionization structure (an equivalent ionized column depth, $N^{*}(N) = \int_{0}^{N} \frac{\gamma}{\Lambda} dN^{\prime}$).  The heating function in this form is then evaluated numerically as a function of time and position, given the beam parameters.

\subsection{Bremsstrahlung emissions}
\label{brems}

An important element of this work is predicting the GOES class of our model flares and so we must evaluate both thermal and non-thermal bremsstrahlung as a function of time.  We note the following important properties of GOES relevant in evaluating this emission.  (1) GOES observes in the bands 1-8 \AA\ (1.55 to 12.41 keV) and 0.5-4 \AA\ (3.10 to 24.82 keV), with flares being classified by the peak flux in the former.  (2) GOES light curves are not spatially resolved; {\it i.e.} GOES does not distinguish coronal sources from foot-points. (3) GOES light curves are a summation of thermal and non-thermal bremsstrahlung, as well as background emission.  We evaluate both thermal and non-thermal bremsstrahlung everywhere along the flaring loop as a function of time and sum the emission to construct a light curve in the native units of GOES (W m$^{-2}$).  

To evaluate the thermal bremsstrahlung, we use Equations 1, 2, and 3 of \citet{culhane1970}, derived for solar flares as observed at terrestrial distance.  These equations include contributions from both free-free and free-bound emission.  We approximate the emission measure by $\int_{V} n^2 dV \approx n_{e} n_{H} A \Delta s$, where $A$ is an estimate of the area based on observations and $\Delta s$ is the length of a given loop segment.  Making these changes, we have, in units of photons sec$^{-1}$ cm$^{-2}$ keV$^{-1}$:

\begin{equation}
I_{thermal} = 3.6 \times 10^{-39} \overline{Z^{2}} k_{B}^{0.3} T_{e}^{-0.2} \epsilon^{-1.3}  \exp{\Bigg( \frac{- \epsilon}{k_{B} T_{e}}\Bigg)} \Bigg[1 - \Big(\frac{\epsilon}{88.0}\Big)^{k_{B} T_{e} / 3}\Bigg]^{-1} n_{e} n_{H} A \Delta s
\end{equation}

\noindent where $\overline{Z^{2}} = 1.4$ is the average charge-squared of ions and $\epsilon$ is the emitted photon energy (in keV).  Note that this function is only valid for photon energies $1.5 \leq \epsilon \leq 15$ keV \citep{culhane1970}, which encompasses the full 1-8 \AA\ range, and is only valid for temperatures below about 20 MK.  Outside of this range, we use the following expression, derived by combining Equations 1 and 1a of their paper:

\begin{equation}
I_{thermal} = 3.6 \times 10^{-39} \overline{Z^{2}} T_{e}^{-0.5} \epsilon^{-1.0} \exp{\Bigg( \frac{- \epsilon}{k_{B} T_{e}}\Bigg)} n_{e} n_{H} A \Delta s
\end{equation}
 
\noindent These functions will give the emitted thermal bremsstrahlung at a given spatial location along the loop at a given time.  Since GOES has no spatial resolution, the total emission is found by spatially integrating along the loop and over the time period of interest.  

Non-thermal bremsstrahlung is more difficult to evaluate.  It requires knowledge of the beam electron distribution function and the appropriate cross-section of interactions.  Following either \citet{holman2011} or \citet{kontar2011}, thick-target bremsstrahlung observed at terrestrial distance is given in general by (in photons sec$^{-1}$ cm$^{-2}$ keV$^{-1}$):

\begin{equation}
I_{thick} = \frac{A}{4 \pi R^{2}} \int_{\epsilon}^{\infty} \mathfrak{F}(E_{0},t) \nu(\epsilon, E_{0}) dE_{0}
\label{thicktargetbrem}
\end{equation}

\noindent where $R = 1.496 \times 10^{13}$ cm (1 AU), $A$ is the loop area (cm$^{2}$), $\mathfrak{F}(E_{0},t)$ is the electron flux spectrum of injected electrons (as before, although converted to units of electrons sec$^{-1}$ cm$^{-2}$ keV$^{-1}$), $E_{0}$ is the initial electron energy, and $\nu$ (below) essentially gives the photon yield for a given electron:

\begin{equation}
\nu(\epsilon, E_{0}) = \int_{E_{0}}^{\epsilon} \frac{ n_{H} v Q({\epsilon, E}) dE}{dE/dt}
\label{nu}
\end{equation}

\noindent for $v$ the electron velocity, $Q({\epsilon, E})$ the cross-section of the interaction (cm$^{2}$ keV$^{-1}$), and $\frac{dE}{dt}$ the energy lost by the electron per unit time (keV sec$^{-1}$).  Note that $\nu$ has units of photons electron$^{-1}$ keV$^{-1}$.  Following \citet{holman2011}, we take

\begin{equation}
\frac{dE}{dt} = n_{H} v \frac{dE}{dN} = n_{H} v \Bigg( \frac{-2 \pi e^{4}}{E} \Big[x \Lambda + (1-x) \Lambda^{\prime}\Big] \Bigg)
\label{dedt}
\end{equation}

\noindent where we now evaluate over the column density, allowing for non-uniform ionization (with variables as defined in previous sections).  Note that for unit consistency, $e^4$ must be in units of cm$^{2}$ keV$^{2}$ with $E$ in keV.

We can reverse the order of integration in Equation \ref{thicktargetbrem}:

\begin{equation}
I_{thick} = \frac{A}{4 \pi R^{2}} \int_{\epsilon}^{\infty} \frac{ n_{H} v Q({\epsilon, E}) dE}{dE/dt} \int_{E}^{\infty} \mathfrak{F}(E_{0},t) dE_{0}
\label{reversebrem}
\end{equation}

\noindent where the inside integral is now analytic, allowing for easier numerical integration regardless of what form of cross-section is chosen.  Evaluation of this integral is explained in detail in Appendix \ref{reverse}.

We must now decide what cross-section is appropriate for non-thermal bremsstrahlung emissions.  At the energies we are considering (1.55 to 12.41 keV and 3.10 to 24.82 keV), an appropriate choice is the Bethe-Heitler cross-section \citep{bethe1934, koch1959, johns1992a, johns1992b}, with the Elwert correction factor \citep{elwert1939}, given by:

\begin{eqnarray}
Q(\epsilon, E) &=& \frac{16 \overline{Z^{2}} r_{0}^{2} \alpha}{3} \frac{m_{e}^{2} c^{4}}{\epsilon E (E + 2m_{e}c^{2})} \ln{\Bigg[ \frac{1 + \Big( \frac{(E-\epsilon)(E - \epsilon + 2m_{e}c^{2})}{E (E+2m_{e}c^{2})}  \Big)^{1/2}}{1 - \Big( \frac{(E-\epsilon)(E - \epsilon + 2m_{e}c^{2})}{E (E+2m_{e}c^{2})}  \Big)^{1/2}} \Bigg]}   \\
&& \times \frac{[E(E+2m_{e}c^{2})]^{1/2} [E - \epsilon + m_{e}c^{2}] \Big[1 - \exp{\Big( - \frac{2 \pi \alpha [E + m_{e}c^{2}]}{[E(E+2m_{e}c^{2})]^{1/2}} \Big)}  \Big] }  {[(E-\epsilon)(E-\epsilon+2m_{e}c^{2})]^{1/2} [E + m_{e}c^{2}] \Big[1 - \exp{\Big( - \frac{2 \pi \alpha [E - \epsilon + m_{e}c^{2}]}{[(E - \epsilon)(E - \epsilon +2m_{e}c^{2})]^{1/2}} \Big)}  \Big] } \nonumber
\label{cross}
\end{eqnarray}

\noindent with $\alpha$ the fine-structure constant and $r_{0}$ the classical electron radius.  We can then numerically integrate the thick-target emissions with the use of Gauss-Laguerre and Gauss-Legendre quadrature, as appropriate.

To facilitate comparisons to observational data, the response of the GOES instrument must be accounted for in the bremsstrahlung calculations.  \citet{hanser1996} describe the response of the X-ray sensor (XRS) aboard GOES-8 in depth.  Note that the sensitivity of the GOES XRS has remained nearly constant for each spacecraft.  We convolve the response function (given by Equation 1 and Figure 3 of \citealt{hanser1996}) with all bremsstrahlung calculations to improve our estimates of GOES classification for the flares studied in this paper.

\section{Results}
\label{results}

We performed 60 numerical experiments exploring the correlation of the GOES classification to the beam properties.  We study two flares observed with RHESSI and base our simulations on their parameters.  The results are split into two sets.  The first set is based on beam parameters for the 2002-04-15 M1.2 flare (Table \ref{20020415sim}) derived by \citet{sui2005}, with a cut-off energy $E_{c} = 24$ keV, a spectral index $\delta(t)$ ranging between about 6 and 10, and a beam power ranging from approximately $6 \times 10^{25}$ to $5 \times 10^{27}$ erg sec$^{-1}$ (see Figure 6 of their paper).  The second set is based on 2002-07-23 X4.8 flare (Table \ref{20020723sim}) derived by \citet{holman2003}, with a cut-off $E_{c}(t)$ ranging from about 18 to 43 keV, spectral index $\delta(t)$ ranging from about 2.5 to 8, and a beam power between $7 \times 10^{26}$ and $2 \times 10^{29}$ erg sec$^{-1}$ (see Figure 3 of their paper).  In both cases, the beam parameters were derived from RHESSI observations of the flares.  The geometry of the 2002-04-15 flare loop (that is, length and cross-sectional area) is based on the estimate of \citet{veronig2004}.  The length of the 2002-07-23 flare loop is estimated from the foot-point locations in \citet{emslie2003}; the area is estimated from \citet{holman2003}, decreased slightly to $7.0 \times 10^{18}$ cm$^{2}$ to calibrate the base run (\# 31) to the observed GOES classification.

In each group of simulations, we vary one beam parameter per simulation in the following way (holding all other values constant).  First, we multiply the beam flux $F_{0}(t)$ by a factor of [$\frac{1}{10},\frac{1}{5},\frac{1}{3},\frac{1}{2},2,3,5,10$].  Second, we change the spectral index of the electron distribution $\delta(t)$ by [$-3,-2,-1,-\frac{1}{2},\frac{1}{2},1,2,3$] (although we limit $\delta$ to 2.01 since it must be strictly greater than 2).  Finally, we change the cut-off energy $E_{c}(t)$ by [$-20,-15,-10,-5,5,10,15,20,30,40,50,75,100$] keV (limiting $E_{c}$ to a minimum of 1 keV).  

Table \ref{20020415sim} summarizes the results for the set of simulations concerning the 2002-04-15 flare.  Table \ref{20020723sim} summarizes the results for the set of simulations concerning the 2002-07-23 flare.  In both cases, we list the run number along with the changes in the beam parameters, the GOES class (in both channels), the maximum temperature (in MK), and the maximum apex density (cm$^{-3}$).  

\begin{longtable}{c c c c c c c c}
\caption{The results of simulations for the 2002-04-15 M1.2 flare.  Run 1 uses beam parameters taken from \citet{sui2005}, while each other simulation modifies one beam parameter, as noted. } \\
\hline
Run \# & Flux & $\Delta \delta$ & $\Delta E_{c}$ & GOES Class & GOES & $T_{\mbox{max}}$ & $n_{\mbox{apex,max}}$ \\
 & & & (keV) & (1-8 \AA) & (0.5-4 \AA) & (MK) & (cm$^{-3}$) \\
\hline
\endfirsthead
1 & 100\% & +0 & +0 & M1.1 & C$3.1$ & 27.3 & $6.3 \times 10^{10}$ \\
2 & 10\% & +0 & +0 & B5.1 & B$3.4$ & 5.15 & $2.1 \times 10^{9}$  \\
3 & 20\% & +0 & +0 & B9.8 & B$6.7$ & 8.20 & $5.6 \times 10^{9}$  \\ 
4 & 33\% & +0 & +0 & C1.6 & C$1.1$ & 12.2 & $1.4 \times 10^{10}$  \\
5 & 50\% & +0 & +0 & C2.5 & C$1.7$ & 17.2 & $2.7 \times 10^{10}$  \\ 
6 & 200\% & +0 & +0 & M4.5 & M$1.1$ & 36.6 & $1.1 \times 10^{11}$  \\ 
7 & 300\% & +0 & +0 & M9.5 & M$2.5$ & 42.6 & $1.5 \times 10^{11}$  \\
8 & 500\% & +0 & +0 & X2.4 & M$7.1$ & 50.1 & $2.1 \times 10^{11}$  \\
9 & 1000\% & +0 & +0 & X10 & X$3.4$ & 62.3 & $3.9 \times 10^{11}$ \\
10 & 100\% & -3 & +0 & C5.6 & C$4.8$ & 22.2 & $4.6 \times 10^{10}$  \\
11 & 100\% & -2 & +0 & C7.4 & C$3.8$ & 24.5 & $5.5 \times 10^{10}$  \\
12 & 100\% & -1 & +0 & C9.5 & C$3.3$ & 26.3 & $6.1 \times 10^{10}$  \\
13 & 100\% & $-\frac{1}{2}$ & +0 & C9.1 & C$3.3$ & 26.9 & $6.0 \times 10^{10}$  \\
14 & 100\% & $+\frac{1}{2}$ & +0 & M1.1 & C$3.1$ & 27.9 & $6.4 \times 10^{10}$  \\
15 & 100\% & +1 & +0 & M1.2 & C$3.2$ & 28.0 & $6.5 \times 10^{10}$  \\
16 & 100\% & +2 & +0 & M1.0 & C$3.0$ & 27.9 & $6.3 \times 10^{10}$ \\
17 & 100\% & +3 & +0 & M1.1 & C$2.9$ & 28.1 & $6.3 \times 10^{10}$  \\
18 & 100\% & +0 & -20 & C7.8 & C$1.3$ & 32.2 & $5.7 \times 10^{10}$  \\
19 & 100\% & +0 & -15 & M1.2 & C$2.5$ & 32.5 & $6.8 \times 10^{10}$  \\
20 & 100\% & +0 & -10 & M1.4 & C$2.9$ & 32.1 & $7.1 \times 10^{10}$  \\
21 & 100\% & +0 & -5 & M1.3 & C$3.1$ & 30.3 & $6.9 \times 10^{10}$  \\
22 & 100\% & +0 & +5 & C5.3 & C$3.8$ & 22.5 & $4.8 \times 10^{10}$  \\
23 & 100\% & +0 & +10 & C5.7 & C$4.6$ & 17.0 & $3.2 \times 10^{10}$  \\
24 & 100\% & +0 & +15 & C6.0 & C$5.1$ & 13.2 & $1.6 \times 10^{10}$  \\
25 & 100\% & +0 & +20 & C6.3 & C$5.6$ & 10.8 & $1.1 \times 10^{10}$  \\
26 & 100\% & +0 & +30 & C6.9 & C$6.5$ & 8.13 & $5.8 \times 10^{9}$  \\
27 & 100\% & +0 & +40 & C7.8 & C$7.6$ & 6.51 & $3.4 \times 10^{9}$ \\
28 & 100\% & +0 & +50 & C7.8 & C$7.9$ & 5.45 & $2.5 \times 10^{9}$  \\
29 & 100\% & +0 & +75 & C8.2 & C$8.6$ & 3.90 & $1.2 \times 10^{9}$  \\
30 & 100\% & +0 & +100 & C8.3 & C$9.1$ & 3.09 & $9.0 \times 10^{8}$  \\
\label{20020415sim}
\end{longtable}

\begin{longtable}{c c c c c c c c}
\caption{The results of simulations for the 2002-07-23 X4.8 flare.  Run 31 uses beam parameters taken from \citet{holman2003}, while each other simulation modifies one beam parameter, as noted.} \\
\hline
Run \# & Flux & $\Delta \delta$ & $\Delta E_{c}$ & GOES Class & GOES & $T_{\mbox{max}}$ & $n_{\mbox{apex,max}}$ \\
 & & & (keV) & (1-8 \AA) & (0.5-4 \AA) & (MK) & (cm$^{-3}$) \\
\hline
\endfirsthead
31 & 100\% & +0 & +0 & X3.6 & X$1.1$ & 28.0 & $1.4 \times 10^{11}$  \\
32 & 10\% & +0 & +0 & M1.2 & C$6.9$ & 6.44 & $6.0 \times 10^{9}$  \\
33 & 20\% & +0 & +0 & M2.3 & M$1.3$ & 11.3 & $1.8 \times 10^{10}$  \\
34 & 33\% & +0 & +0 & M4.7 & M$2.1$ & 17.3 & $4.7 \times 10^{10}$  \\
35 & 50\% & +0 & +0 & M9.3 & M$3.5$ & 21.7 & $7.5 \times 10^{10}$  \\
36 & 200\% & +0 & +0 & X14 & X$5.1$ & 35.7 & $2.3 \times 10^{11}$  \\
37 & 300\% & +0 & +0 & X32 & X$8.4$ & 40.6 & $3.2 \times 10^{11}$  \\
38 & 500\% & +0 & +0 & X80 & X$21$ & 47.3 & $4.7 \times 10^{11}$  \\
39 & 1000\% & +0 & +0 & X300 & X$89$ & 58.3 & $8.4 \times 10^{11}$  \\
40 & 100\% & -3 & +0 & X2.0 & X$1.2$ & 23.8 & $8.1 \times 10^{10}$  \\
41 & 100\% & -2 & +0 & X2.7 & X$1.1$ & 26.4 & $1.1 \times 10^{11}$  \\
42 & 100\% & -1 & +0 & X3.2 & X$1.1$ & 27.3 & $1.2 \times 10^{11}$  \\
43 & 100\% & $-\frac{1}{2}$ & +0 & X3.4 & X$1.0$ & 27.6 & $1.3 \times 10^{11}$  \\
44 & 100\% & $+\frac{1}{2}$ & +0 & X3.6 & X$1.1$ & 28.2 & $1.4 \times 10^{11}$  \\
45 & 100\% & +1 & +0 & X3.5 & X$1.0$ & 28.2 & $1.4 \times 10^{11}$  \\
46 & 100\% & +2 & +0  & X3.8 & X$1.0$ & 28.5 & $1.5 \times 10^{11}$  \\
47 & 100\% & +3 & +0 & X3.7 & X$1.0$ & 28.5 & $1.4 \times 10^{11}$  \\
48 & 100\% & +0 & -20 & M6.1 & M$2.7$ & 18.1 & $6.4 \times 10^{11}$  \\
49 & 100\% & +0 & -15 & X2.6 & M$5.3$ & 27.2 & $1.2 \times 10^{11}$  \\
50 & 100\% & +0 & -10 & X4.0 & M$8.7$ & 29.9 & $1.5 \times 10^{11}$  \\
51 & 100\% & +0 & -5 & X3.9 & M$9.9$ & 29.5 & $1.5 \times 10^{11}$  \\
52 & 100\% & +0 & +5 & X2.8 & X$1.0$ & 26.5 & $1.2 \times 10^{11}$  \\
53 & 100\% & +0 & +10 & X1.5 & X$1.1$ & 28.2 & $4.1 \times 10^{10}$  \\
54 & 100\% & +0 & +15 & X1.6 & X$1.2$ & 15.7 & $4.3 \times 10^{10}$ \\
55 & 100\% & +0 & +20 & X1.7 & X$1.4$ & 11.9 & $2.4 \times 10^{10}$  \\
56 & 100\% & +0 & +30 & X1.8 & X$1.7$ & 8.26 & $1.1 \times 10^{10}$ \\
57 & 100\% & +0 & +40 & X2.0 & X$1.9$ & 6.18 & $6.4 \times 10^{9}$  \\
58 & 100\% & +0 & +50 & X2.2 & X$2.1$ & 5.04 & $4.3 \times 10^{9}$  \\
59 & 100\% & +0 & +75 & X2.3 & X$2.4$ & 3.44 & $2.2 \times 10^{9}$  \\
60 & 100\% & +0 & +100 & X2.4 & X$2.6$ & 2.64 & $1.5 \times 10^{9}$  \\
\label{20020723sim}
\end{longtable}

\begin{figure}
\begin{minipage}[b]{0.5\linewidth}
\centering
\includegraphics[width=3in]{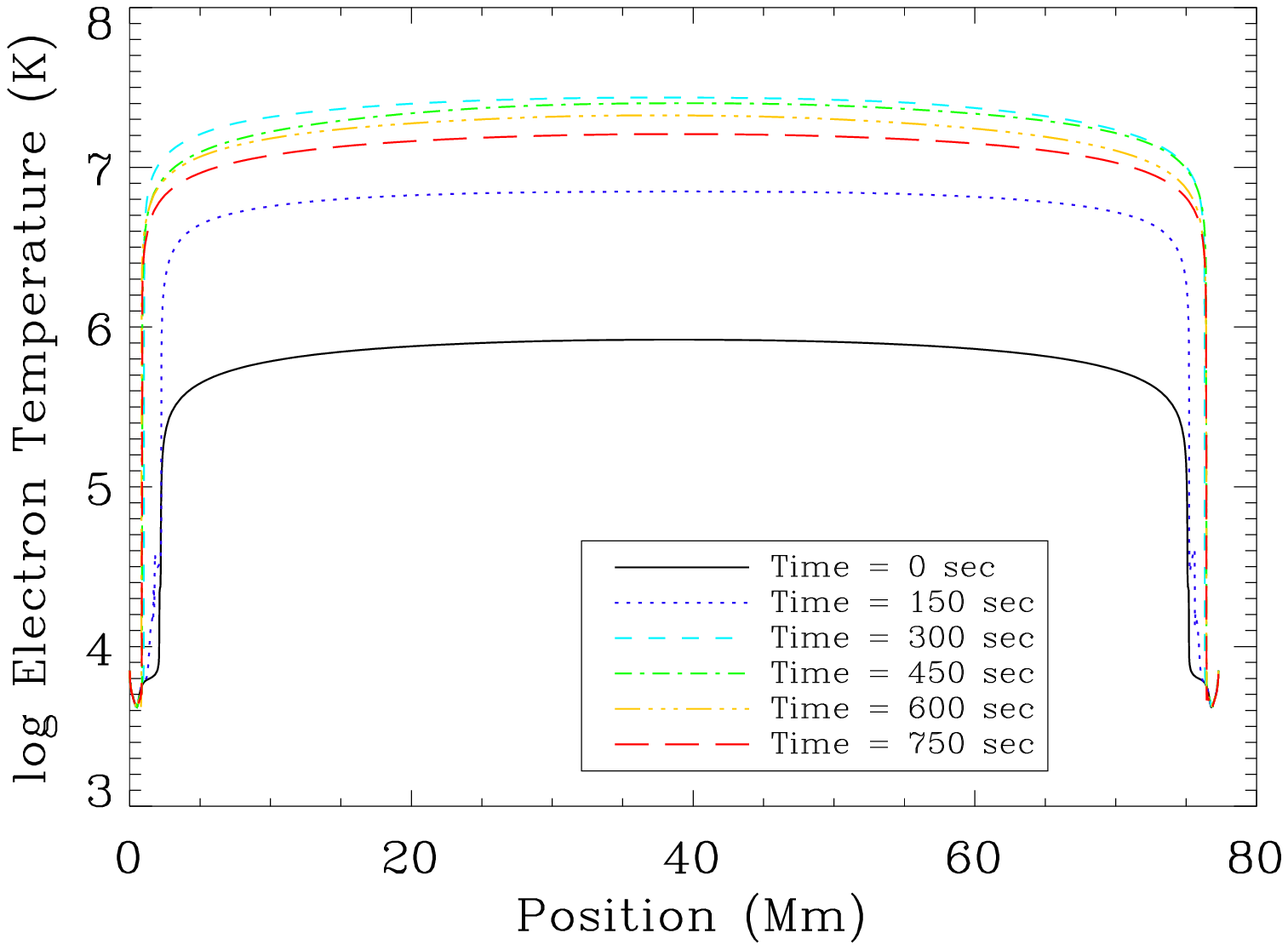}
\end{minipage}
\hspace{0.1in}
\begin{minipage}[b]{0.5\linewidth}
\centering
\includegraphics[width=3in]{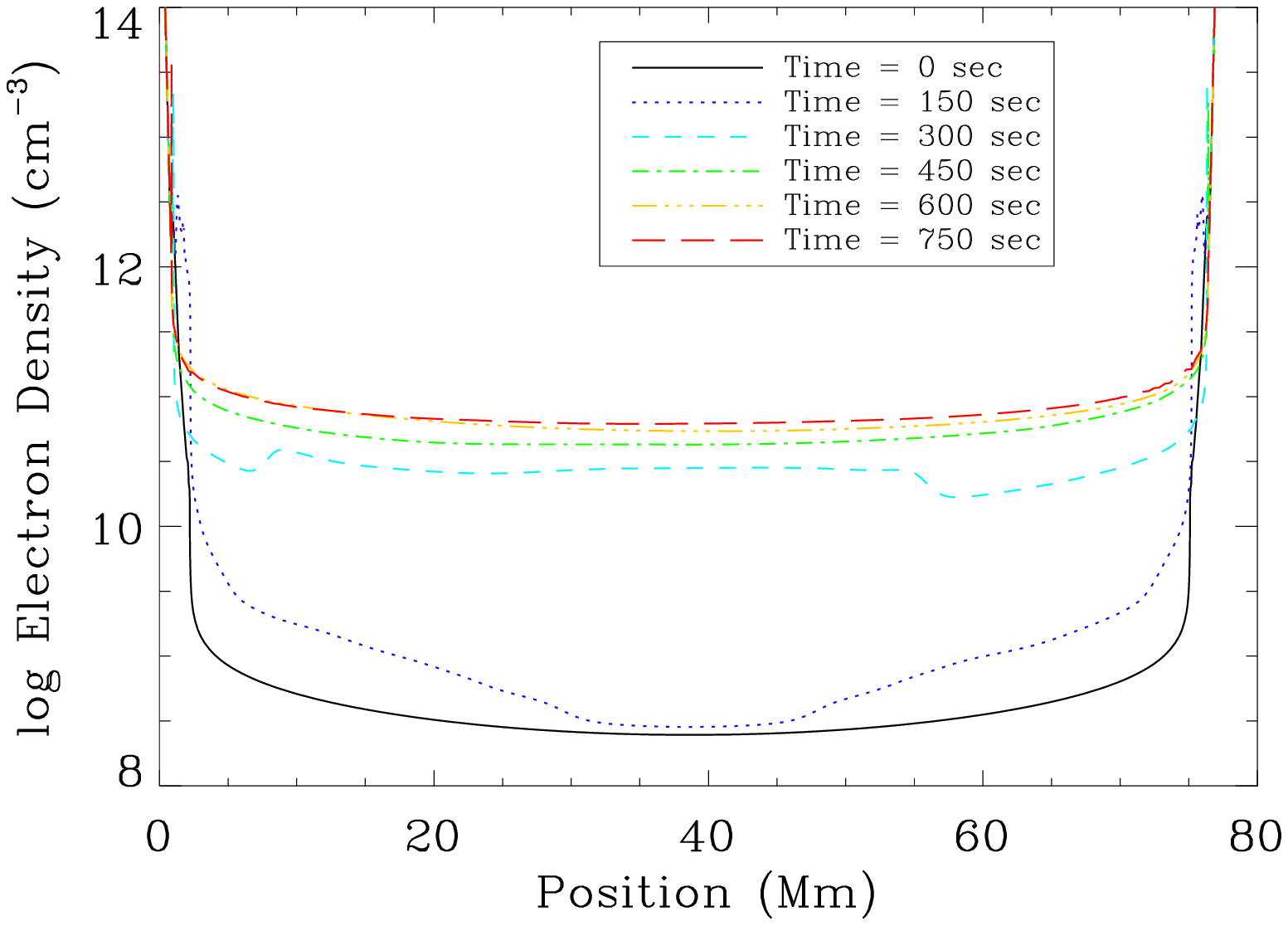}
\end{minipage}
\begin{minipage}[b]{0.5\linewidth}
\centering
\includegraphics[width=3in]{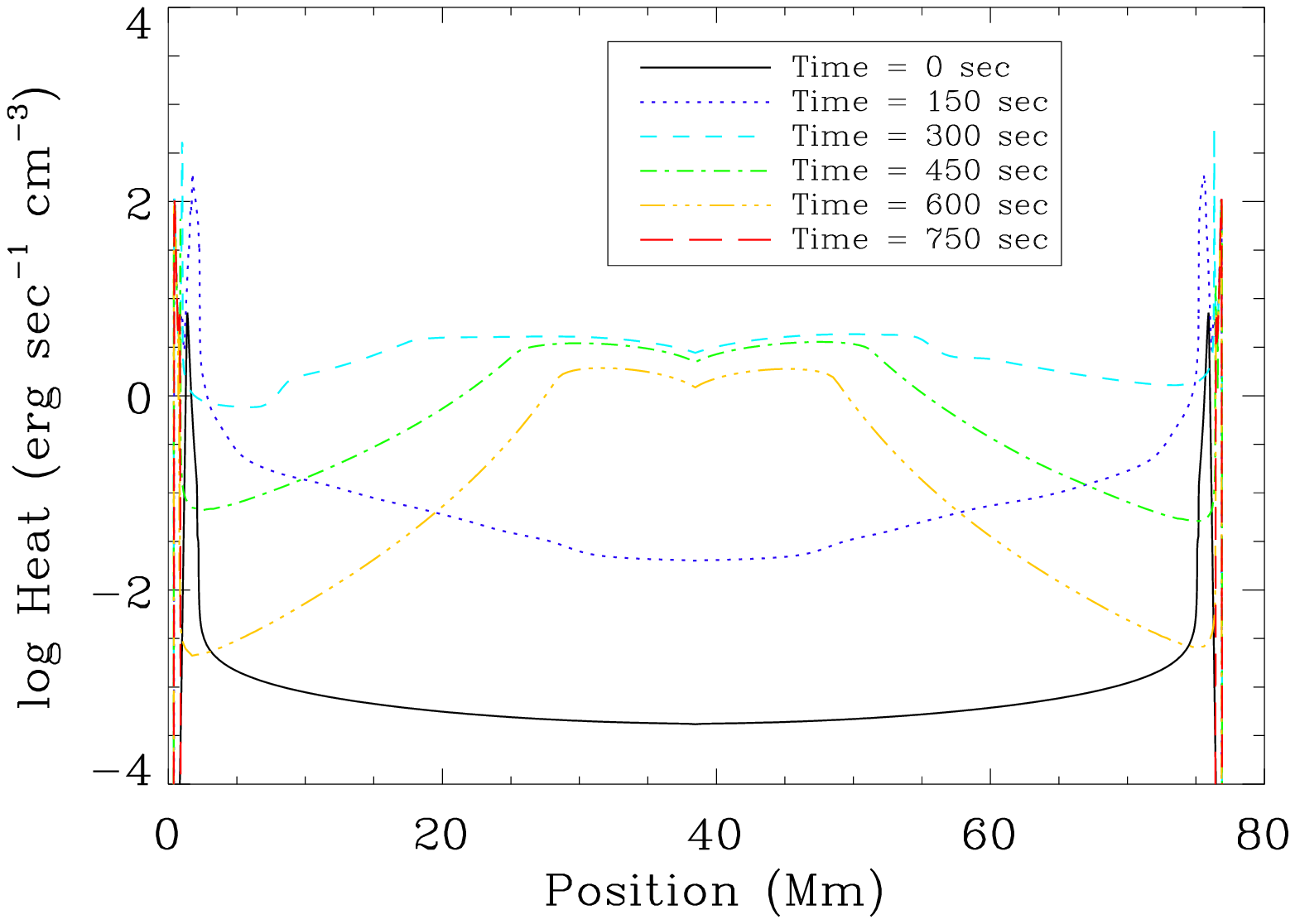}
\end{minipage}
\hspace{0.1in}
\begin{minipage}[b]{0.5\linewidth}
\centering
\includegraphics[width=3in]{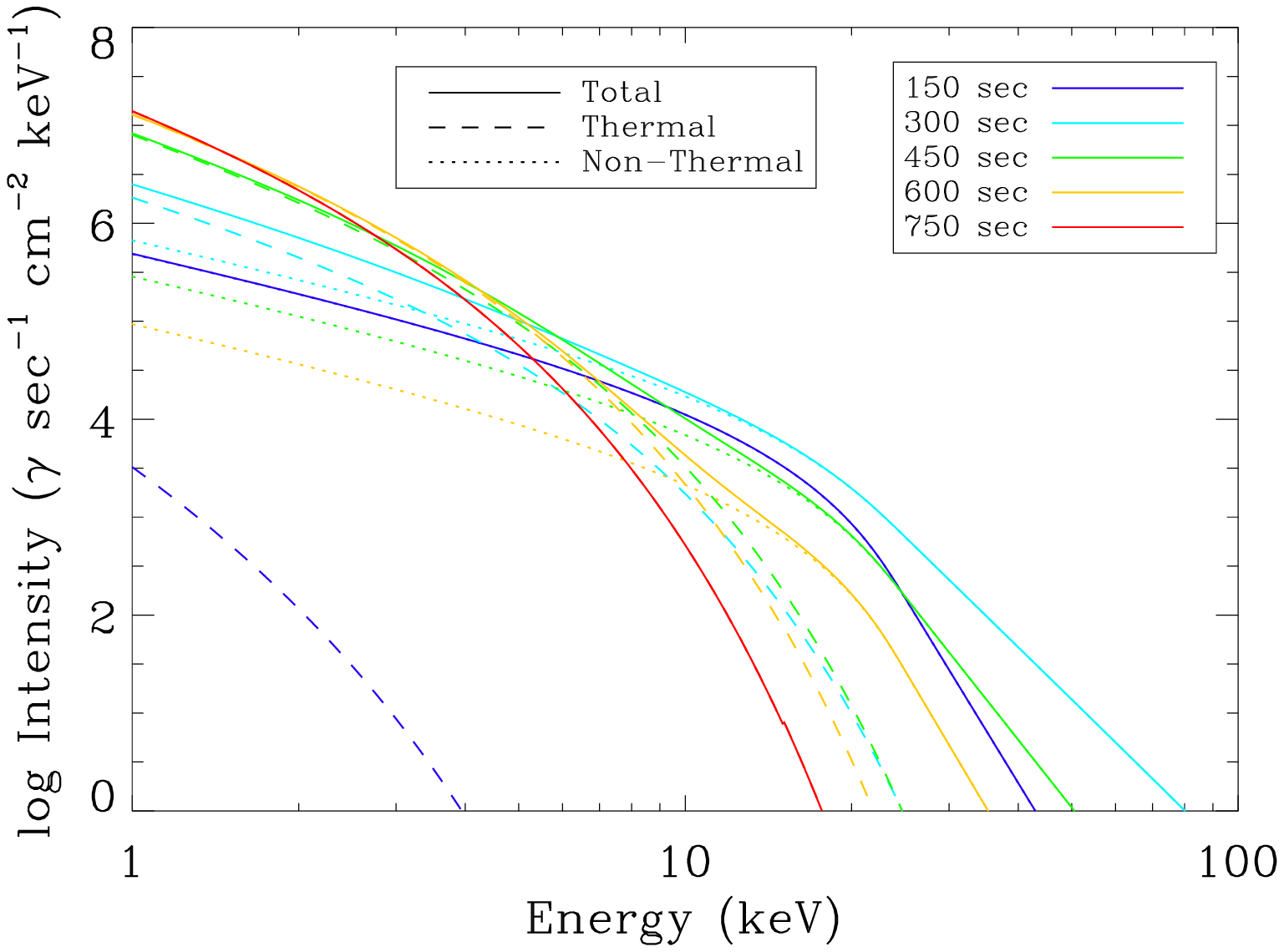}
\end{minipage}
\begin{minipage}[b]{\linewidth}
\centering
\includegraphics[width=3in]{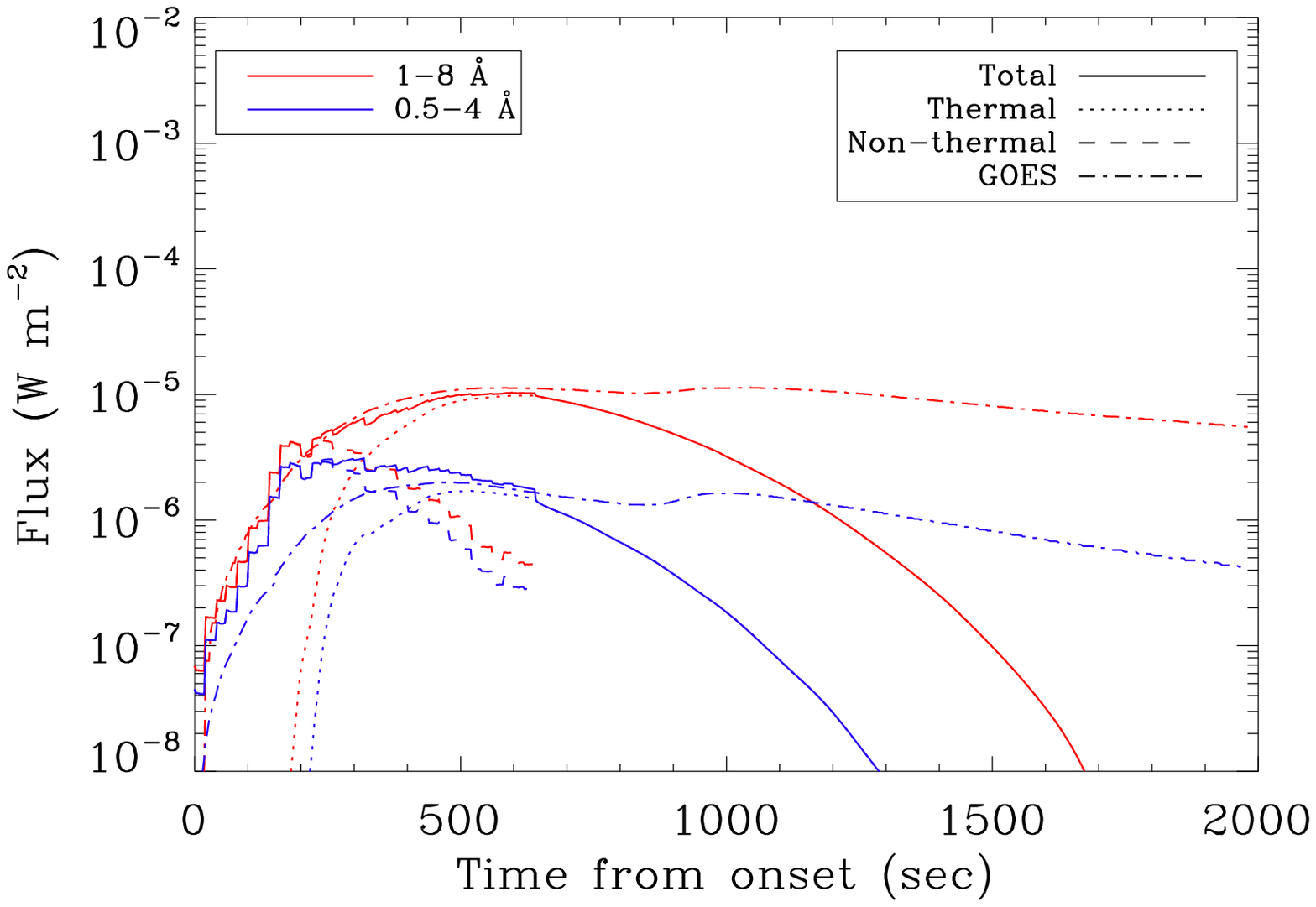}
\end{minipage}
\caption{Results for Run~1, beam parameters as given in Figure 6 of \citet{sui2005}.  {\it Top Left:} Electron temperature vs position, with a select few times times overlaid.  {\it Top Right:} Electron density vs position.  {\it Middle Left:} Heat input vs position.  {\it Middle Right:} Predicted X-ray spectra of the flare, integrated over half of the loop.   {\it Bottom:} Predicted GOES light curve, peaking at $1.1 \times 10^{-5}$ W m$^{-2}$ (M1.1) in the 1-8 \AA\ channel and $3.1 \times 10^{-6}$ W m$^{-2}$ in the 0.5-4 \AA\ channel.  The observed GOES light curves, starting at 15 April 2002 23:07 UT, are overlaid (and have been background subtracted).  }
\label{20020415base}
\end{figure}

For example, consider Figure \ref{20020415base}, which shows the results of Run 1.  The beam lasts 640 seconds, for a loop of length $2L = 77$ Mm.  The upper panels show the electron temperature (reaching a peak of 27.3 MK) and density (with a maximum apex value near $6 \times 10^{10}$ cm$^{-3}$) as functions of position, at different times as indicated.  The temperature rises for around 300 seconds, when the beam begins to weaken.  There is significant radiative cooling in the chromosphere; in the corona, thermal conduction drives energy losses initially, gradually transitioning to radiative cooling as the density rises and temperature falls.  The density in the corona rises due to chromospheric ablation throughout the duration of heating, and the maximum intensity occurs when the heating ceases (a few minutes after it reaches its maximum temperature).  The center left panel similarly shows the heat deposition (which includes background heating).  As the loop fills due to chromospheric ablation, the heat deposition occurs higher and higher in the corona as the mean-free path of beam electrons decreases.  Although the beam continues for 640 seconds, the beam flux begins to fall around 300 seconds, and thus the heat deposition begins to fall.  The center right figure shows the spectrum calculated at one of the footpoints at a few selected times.  Initially ($t = 150$ sec), the emission is entirely non-thermal bremsstrahlung, but as the loop begins to heat and fill, thermal bremsstrahlung eventually becomes the dominant component of the emission at lower energies (e.g., at 600 seconds, the thermal emission at 1 keV is more than 100 times stronger than the non-thermal emission).  Finally, the predicted GOES light curve (with 1-8 \AA\ in red and 0.5-4 \AA\ in blue) is shown at bottom right, with the relative amounts of thermal and non-thermal bremsstrahlung indicated.  The light curve  (1-8 \AA) peaks at $1.1 \times 10^{-5}$ W m$^{-2}$, corresponding to M1.1, while the 0.5-4 \AA\ component peaks at $3.1 \times 10^{-6}$ W m$^{-2}$ (C3.1).  The actual (background subtracted) GOES light curves, starting at 15 April 2002 23:07 UT, are overlaid for comparison (peaking at M1.1 and C2.0, respectively).  The light curve is initially non-thermal bremsstrahlung, but as the loops heats and fills up, thermal bremsstrahlung increases until it becomes the dominant component in the passband.

\begin{figure}
\begin{minipage}[b]{0.5\linewidth}
\centering
\includegraphics[width=3in]{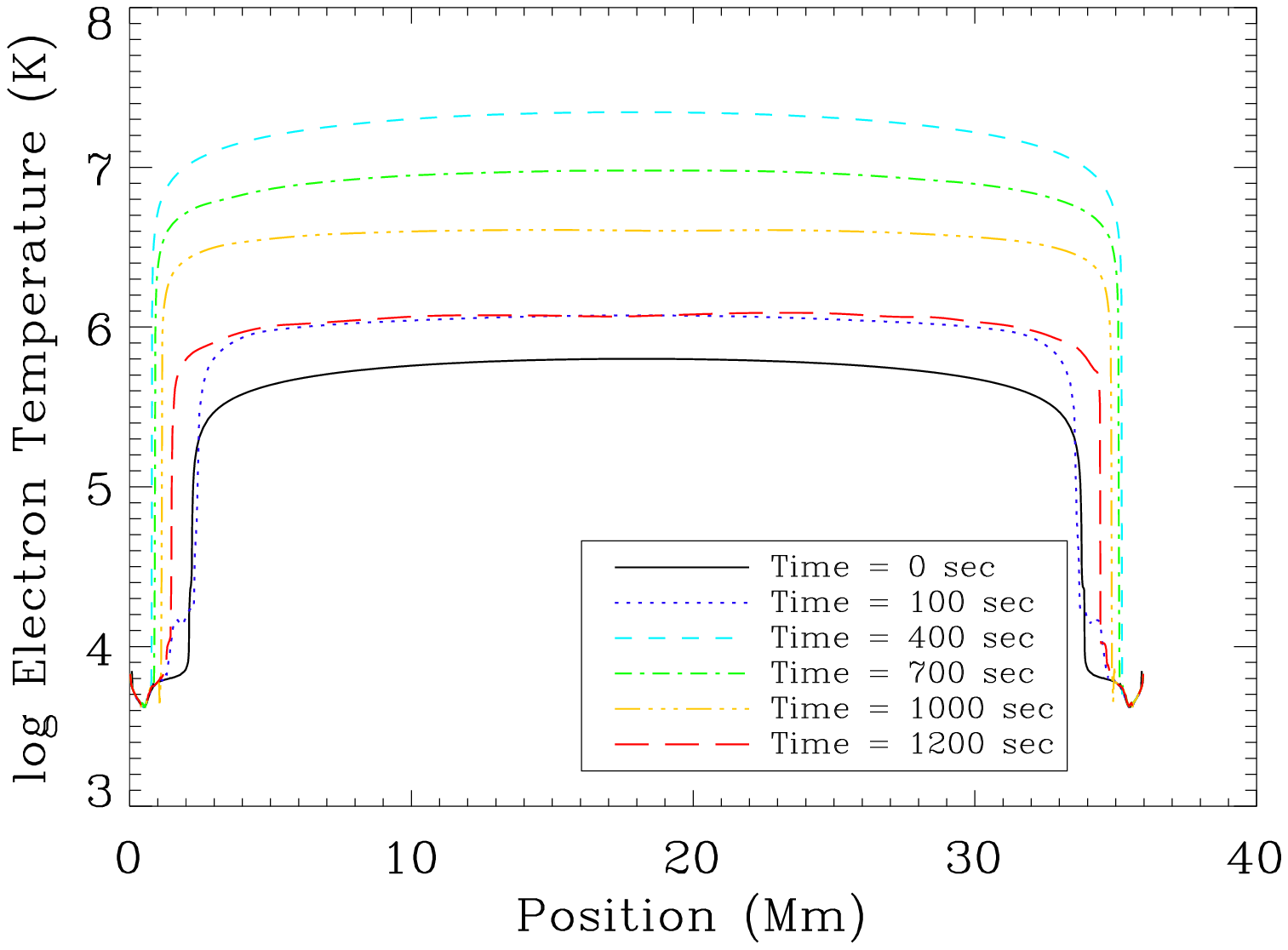}
\end{minipage}
\hspace{0.1in}
\begin{minipage}[b]{0.5\linewidth}
\centering
\includegraphics[width=3in]{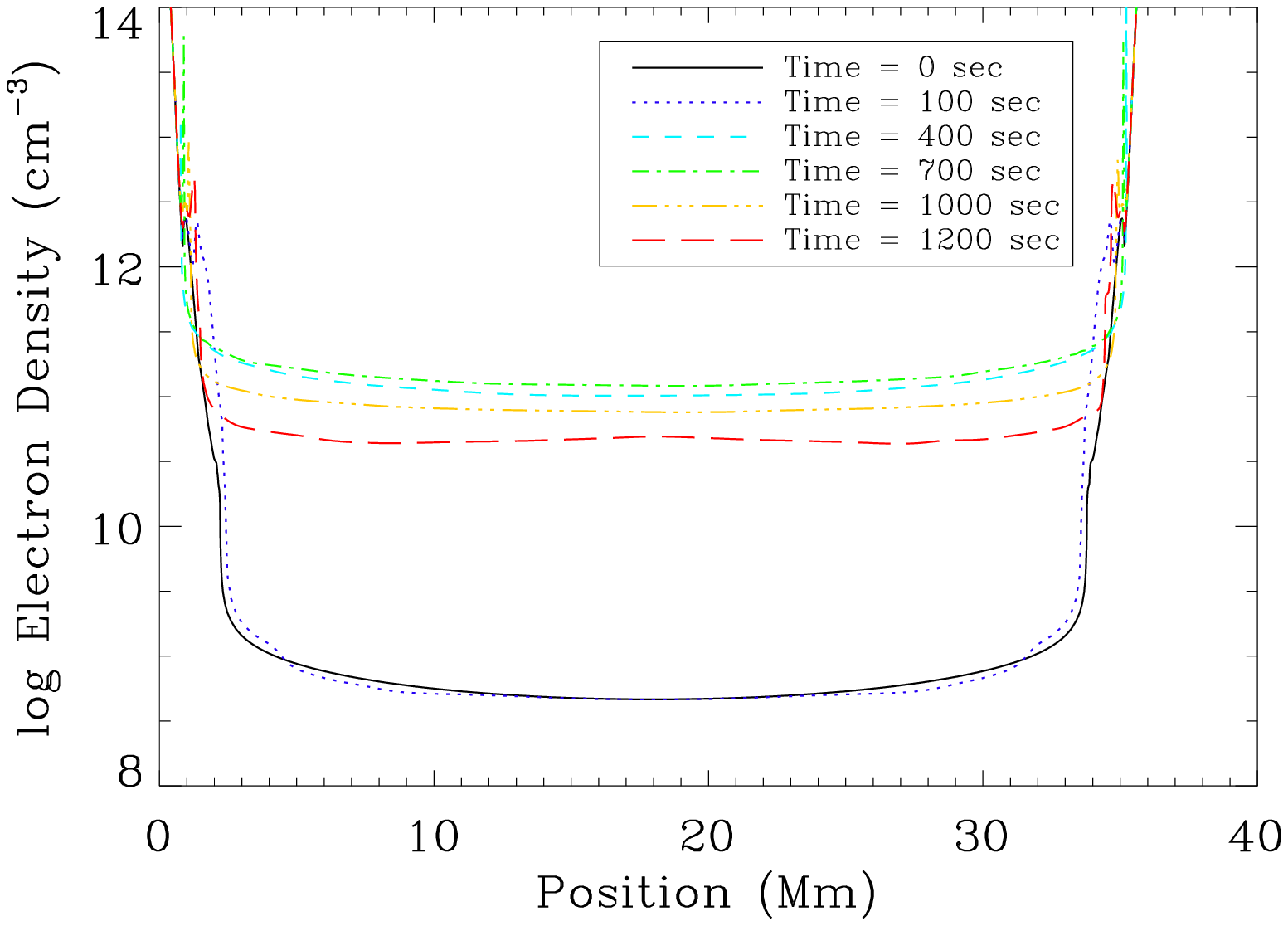}
\end{minipage}
\begin{minipage}[b]{0.5\linewidth}
\centering
\includegraphics[width=3in]{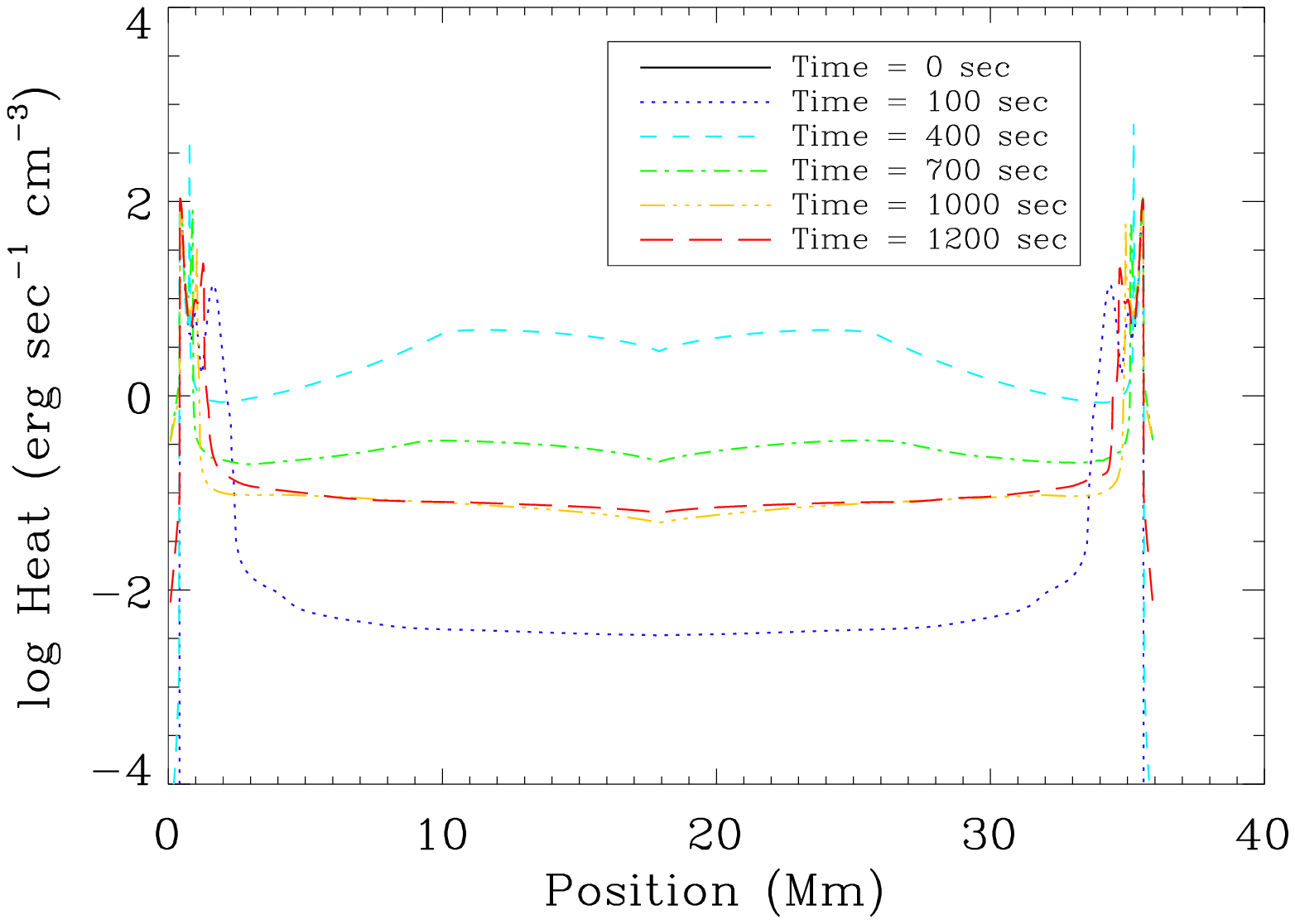}
\end{minipage}
\hspace{0.1in}
\begin{minipage}[b]{0.5\linewidth}
\centering
\includegraphics[width=3in]{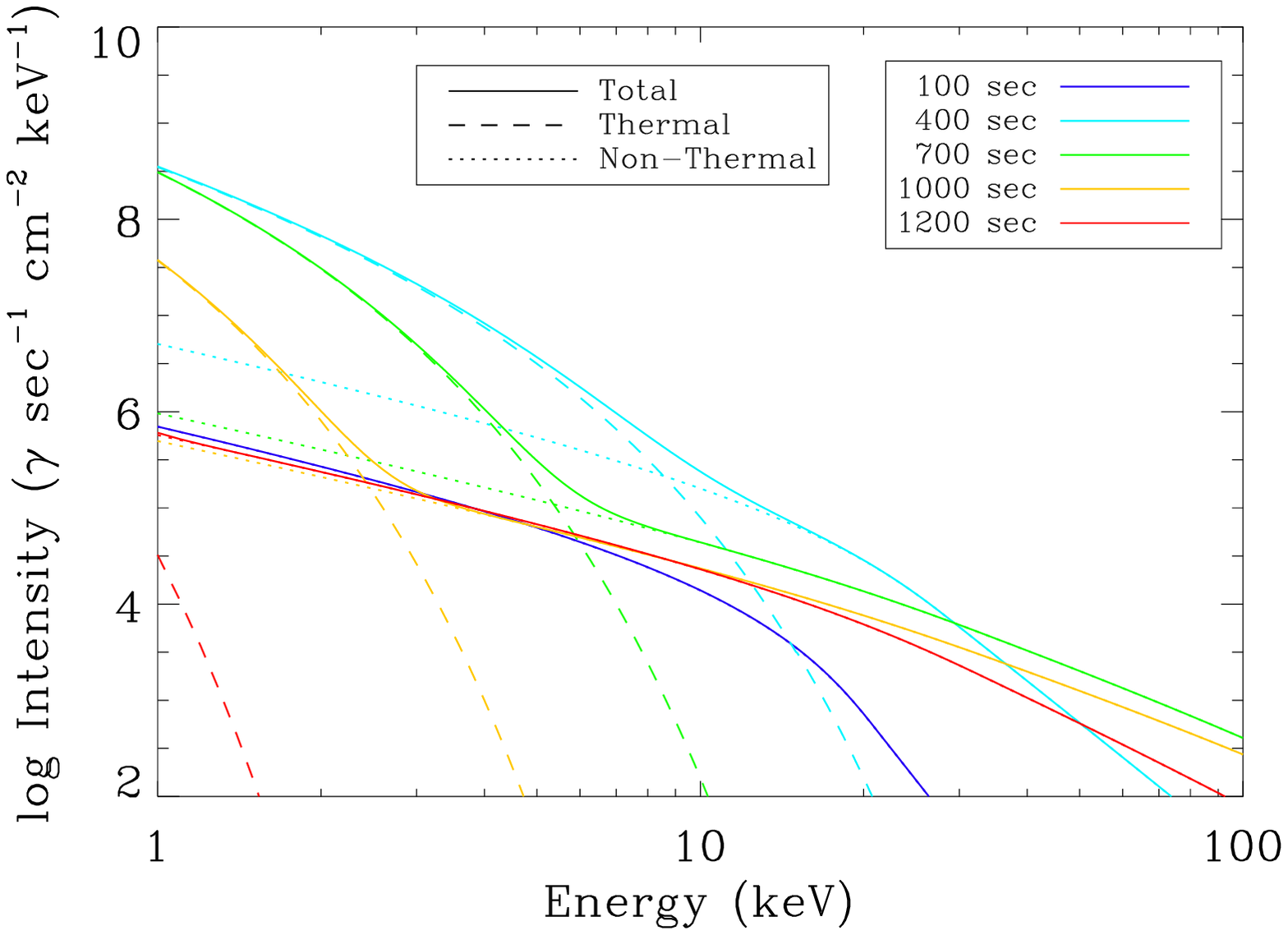}
\end{minipage}
\begin{minipage}[b]{\linewidth}
\centering
\includegraphics[width=3in]{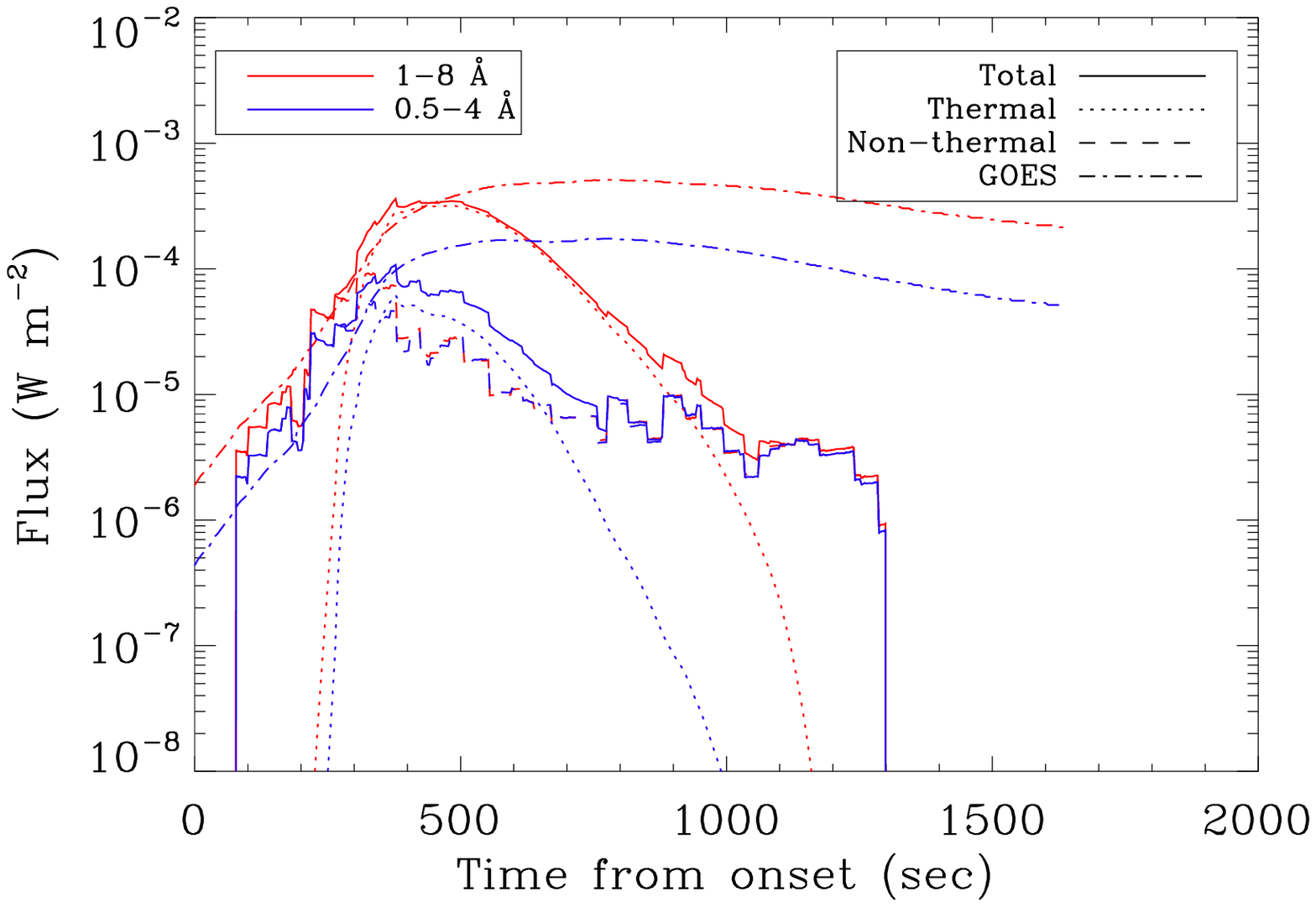}
\end{minipage}
\caption{Results for Run~31, beam parameters as given in Figure 3 of \citet{holman2003}.  {\it Top Left:} Electron temperature vs position, with a select few times times overlaid.  {\it Top Right:} Electron density vs position.  {\it Middle Left:} Heat input vs position.   {\it Middle Right:} Predicted X-ray spectra of the flare, integrated over half of the loop.   {\it Bottom:} Predicted GOES light curve, peaking at $3.6 \times 10^{-4}$ W m$^{-2}$ (X3.6) in the 1-8 \AA\ channel and $1.1 \times 10^{-4}$ W m$^{-2}$ in the 0.5-4 \AA\ channel.  The observed GOES light curves, starting at 23 July 2002 00:22 UT, are overlaid (and have been background subtracted). }
\label{20020723base}
\end{figure}

Similarly, Figure \ref{20020723base} shows the evolution of Run 31.  The beam lasts for 1220 seconds, for a loop of length of $2L = 36$ Mm.  The predicted light curve in this case peaks at $3.6 \times 10^{-4}$ W m$^{-2}$ (X3.6) in the 1-8 \AA\ channel and $1.1 \times 10^{-4}$ W m$^{-2}$ (X1.1) in the 0.5-4 \AA\ channel.  As before, the temperature rises sharply due to the energy deposition, which then triggers chromospheric ablation.  As the loop fills and heats up, the contribution of thermal bremsstrahlung to the light curve rises sharply.  In this case, the loop begins to cool and drain before the beam ceases.  In both cases, the GOES class is in approximate agreement with the observations.  Note that the results are fairly insensitive to the initial density profile.  If we had assumed an initial coronal density of $10^{10}$ cm$^{-3}$, over a loop length of 16 Mm, we would find that only electrons with energy $\lesssim 12$ keV would be stopped in the corona, and thus the majority of energy would still be deposited in the chromosphere (using the estimate of stopping depth from \citealt{nagai1984}).  We now turn our attention towards altering the beam parameters one at a time, to examine their effect on the GOES class.

\begin{figure}
\centering
\includegraphics[width=3.5in]{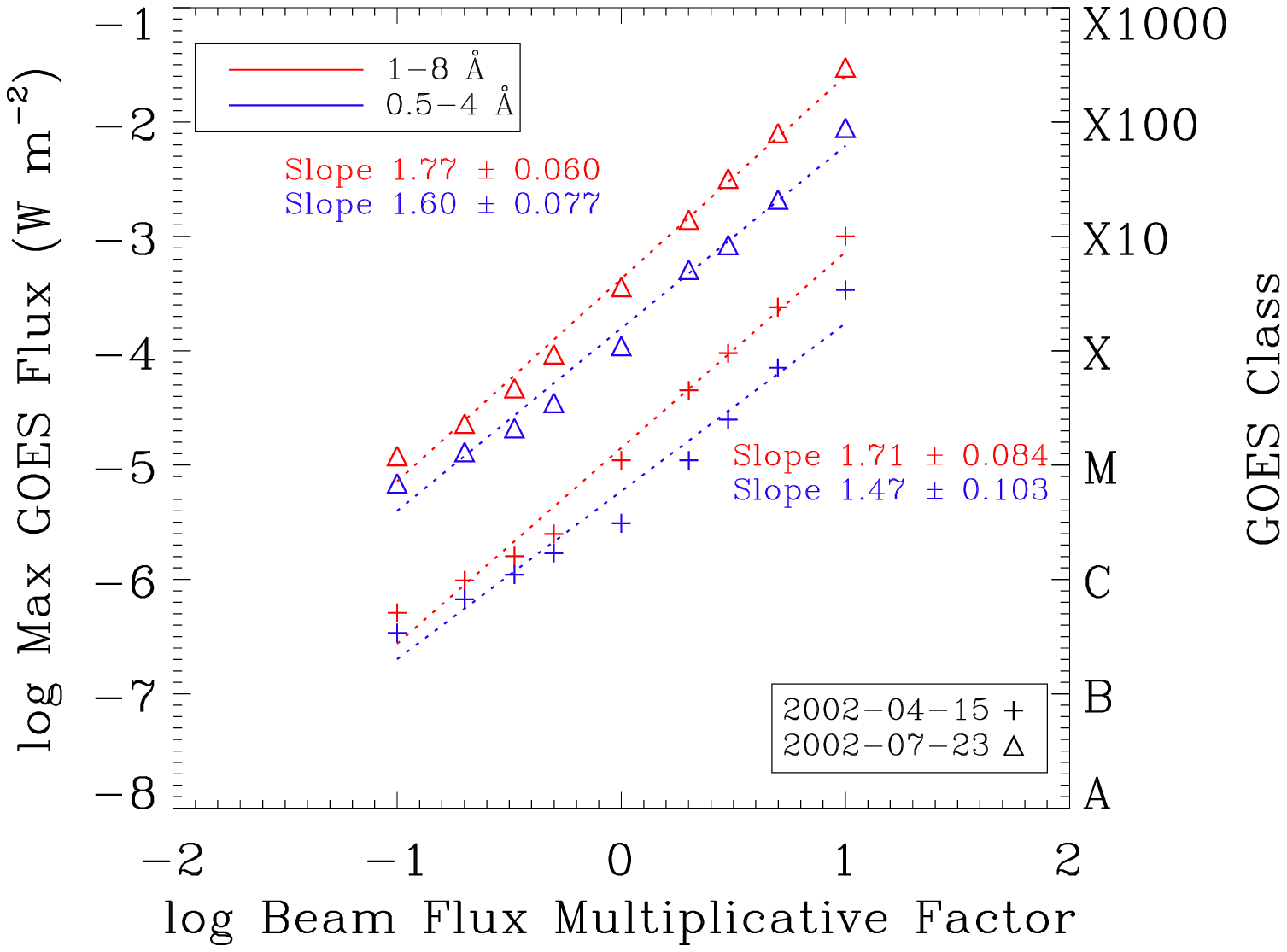}
\includegraphics[width=3.5in]{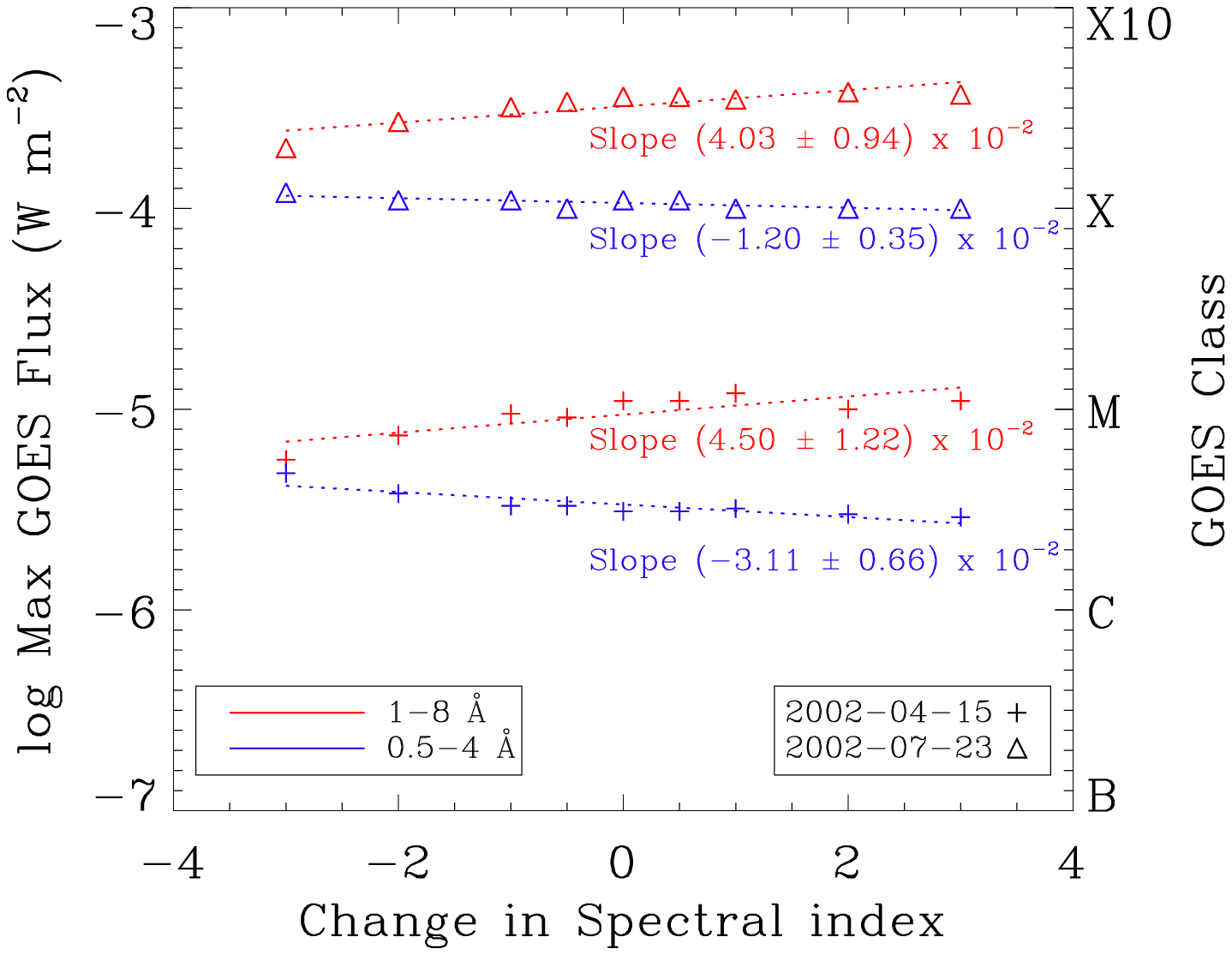}
\includegraphics[width=3.5in]{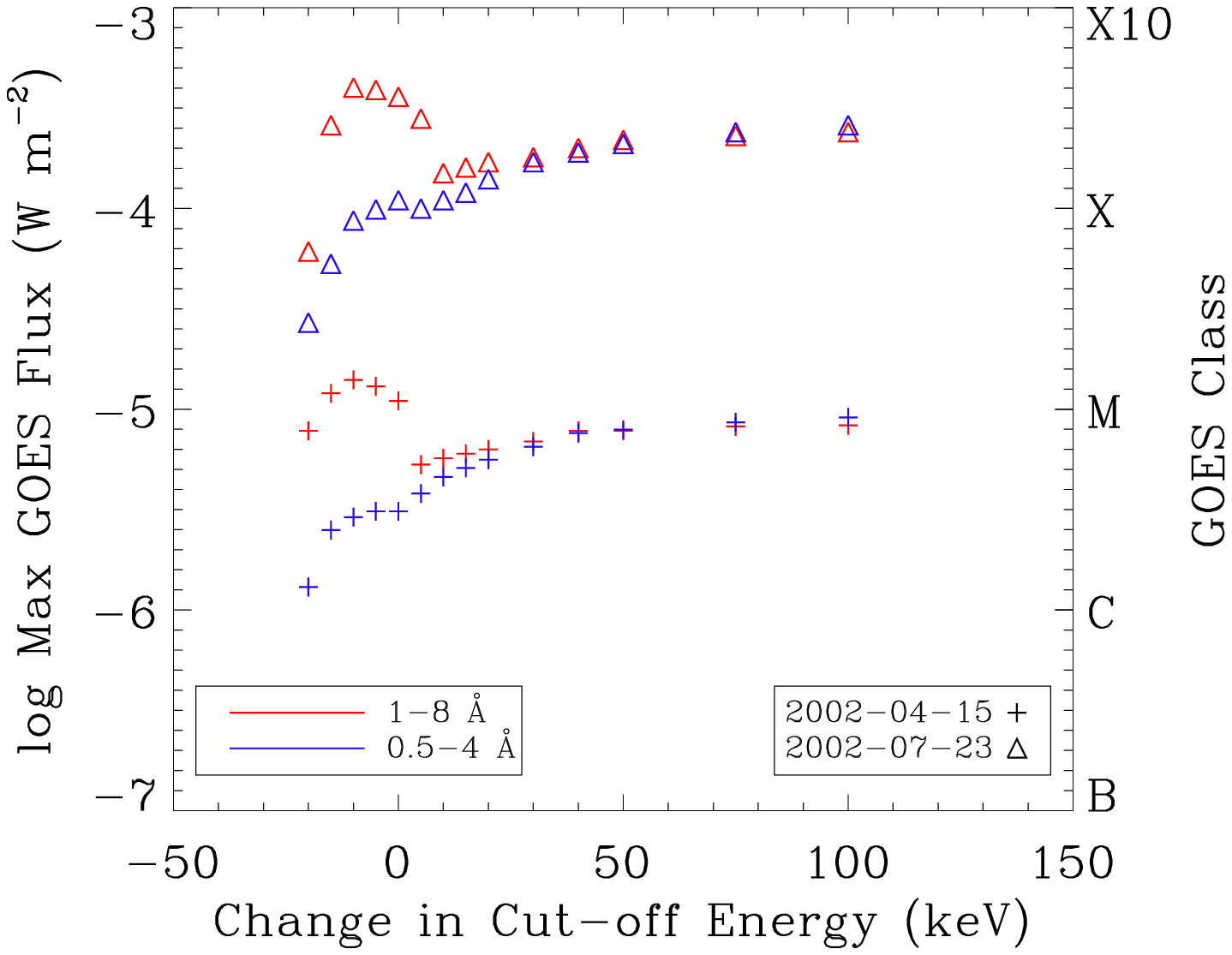}
\caption{The relations between GOES class and beam parameters.  Each simulation (excluding the base case) varies one parameter at a time.  {\it Top:} The correlation between beam flux and GOES class.  {\it Middle:} The correlation between spectral index of beam electrons and GOES class.  {\it Bottom:} The relationship between lower cut-off energy of beam electrons and GOES class.}
\label{goesbeam}
\end{figure}

First, holding all other values constant, we multiply the beam flux by [$\frac{1}{10},\frac{1}{5},\frac{1}{3},\frac{1}{2},2,3,5,10$] for both flares (equivalent to multiplying the total non-thermal energy by the same amount).  There is a clear correlation between the GOES class and the beam flux (see the first plot in figure \ref{goesbeam}).  For Runs 1-9, we find that the GOES class $\Psi$ (in the 1-8 \AA\ channel) is proportional to the beam energy $E$, such that $\Psi \propto E^{\alpha}$ for $\alpha = 1.71 \pm 0.084$; for Runs 31-39, we find $\alpha = 1.77 \pm 0.060$.  In the 0.5-4 \AA\ channel, we find a similar trend: for Runs 1-9, we find $\Psi \propto E^{\alpha}$ for $\alpha = 1.47 \pm 0.103$ and for Runs 31-39 $\alpha = 1.60 \pm 0.077$.  In Runs 2-5 and 32-33 (and 34 in the 0.5-4 \AA\ channel), the loops were not heated enough to produce significant thermal bremsstrahlung, and thus their GOES classes were primarily determined by non-thermal emissions.  Note that non-thermal emissions are linearly proportional to the beam flux (Equation \ref{thicktargetbrem}); if we fit a line to just Runs 2-5, we find an exponent $\alpha = 0.98 \pm .023$ in the 1-8 \AA\ channel and $\alpha = 1.00 \pm 0.014$ in the 0.5-4 \AA\ channel, confirming that the non-thermal emission is proportional to the mean beam energy.  In all the other runs, the peak of the light curves was a combination of thermal and non-thermal emissions, with thermal emissions being indirectly related to the beam flux.  If we remove the simulations without thermal emissions and recalculate, we find exponents $\alpha = 1.94 \pm 0.026$ (1-8 \AA) and $\alpha = 2.04 \pm 0.054$ (0.5-4 \AA) for Runs 1 and 6-9, and $1.91 \pm .014$ (1-8 \AA) for Runs 31 and 34-39 and $1.84 \pm 0.050$ (0.5-4 \AA) for Runs 31 and 35-39.  Thus, the maximum GOES class $\Psi$ is related to the total beam energy $E$, $\Psi \propto E^{\alpha}$ for $\alpha \approx 1.7$ (1-8 \AA) and $\alpha \approx 1.6$ (0.5-4 \AA).  There are two reasons why the values differ slightly for each flare: different loop lengths and different cut-off energies, both of which affect the amount of heating and thus thermal bremsstrahlung produced. 

This result should be compared to that of \citet{warren2004}, who showed that $\Psi \propto E^{1.75}$ (1-8 \AA) and $\Psi \propto E^{2.25}$ (0.5-4 \AA) using simple analytic considerations.  Our results are in agreement for the 1-8 \AA\ channel, while they are in contrast in the higher energy channel.  If the emission were entirely non-thermal, then we would find a linear relationship ($\Psi \propto E$, Equation \ref{thicktargetbrem}), regardless of which channel is being considered.  Since the 0.5-4 \AA\ channel is more sensitive to higher energies, we would expect it to be closer to linear than the lower energy channel as it would be more sensitive to the non-thermal emissions, and thus the exponent $\alpha$ should be lower.  The reason for this contrast is perhaps their use of the hydrostatic scaling laws derived for the corona, whereas the non-thermal emissions will be primarily at chromospheric depths.  It should also be noted that our result, as well as that of \citet{warren2004}, are in contrast to what is expected from the Neupert effect.  As shown by \citet{lee1993,lee1995} by using the Neupert effect relation, the maximum flux in the SXRs should be linearly proportional to the energy deposited by the electron beam.  \citet{warren2004} suggest that this difference is due to the peak SXR flux being dependent upon other flare parameters, to which we now turn our attention.

\begin{figure}
\centering
\includegraphics[width=3.25in]{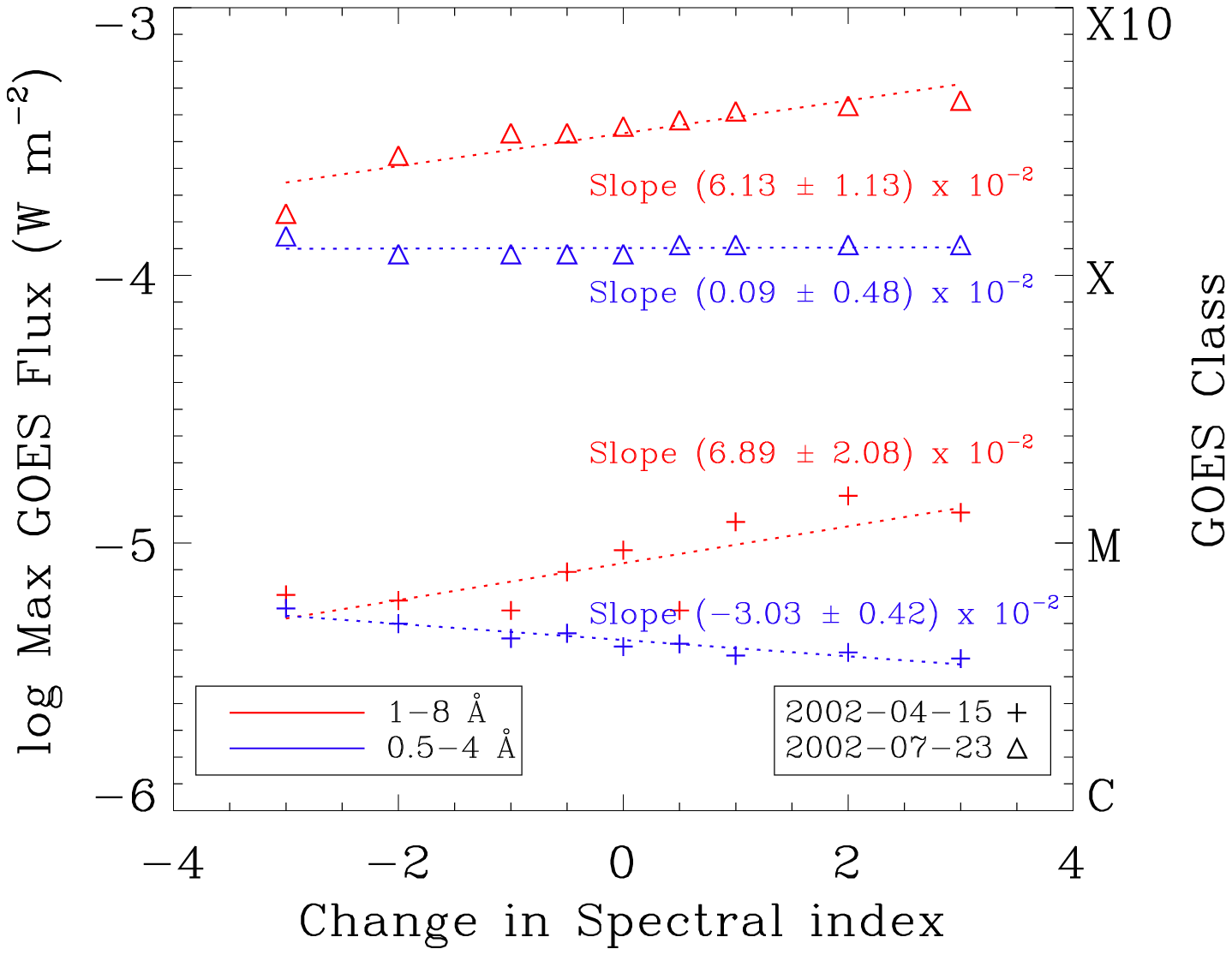}
\caption{The change in the beam index versus maximum GOES flux, using a sharp low-energy cut-off instead of a low-energy knee.  Compare to the second plot in Figure \ref{goesbeam}, which used a low-energy knee.}
\label{indexsharp}
\end{figure}

Next, holding all the other values constant again, we add [$-3,-2,-1,-\frac{1}{2},\frac{1}{2},1,2,3$] to the spectral index $\delta(t)$.  Varying the spectral index of the electron distribution has only a small effect on the GOES class (see the second plot in Figure \ref{goesbeam}).  For Runs 1 and 10-17, with a change of $\delta \pm 3$, the GOES flux (1-8 \AA) varies by a factor of 2, and about a factor of 3 in the higher energy channel; for Runs 31 and 40-47, the flux varies by less than a factor of 2 in both channels.  In all cases, though, an increase in the spectral index does increase the GOES class slightly in the 1-8 \AA\ channel and decreases it slightly in the 0.5-4 \AA\ channel.  The reason is that an increase in the index reduces the proportion of high- to low-energy electrons, so that more energy is deposited higher up in the atmosphere, increasing the temperature and thus the amount of thermal bremsstrahlung, while slightly reducing the non-thermal emissions.  Thus, the channel which is more sensitive to thermal emissions increases slightly, and the channel more sensitive to non-thermal emissions decreases slightly.  The immediate question is whether this result is due to the assumed form of the beam electron distribution ({\it i.e.} the importance of the low-energy knee electrons).  Runs 1, 10-17, 31, and 40-47 were repeated using a sharp low-energy cut-off ($\mathfrak{F}_{0}(E_{0},t) = 0$ for $E_{0} < E_{c}$), but equivalent amount of total beam energy (see Figure \ref{indexsharp}).  Although the slopes are larger in this case, they remain small.  The conclusion is unaltered: the spectral index only marginally affects the GOES class.  

The primary reason that the spectral index minimally affects the GOES flux is because the total beam energy was held constant.  Changing the spectral index is equivalent to slightly altering the proportion of high- to low-energy electrons, which in turn affects the mean location of energy deposition.  Increasing the index decreases the number of high-energy electrons in the beam, which then causes deposition of energy slightly higher up the loop (and vice versa).  Compare \citet{dennis1985}, who found that the photon spectral index $\gamma$ is not correlated to the peak HXR flux (and in turn, $\delta = \gamma + 1$, so that the relation holds for the electron index as well).

\begin{figure}
\begin{minipage}[b]{0.5\linewidth}
\centering
\includegraphics[width=3.25in]{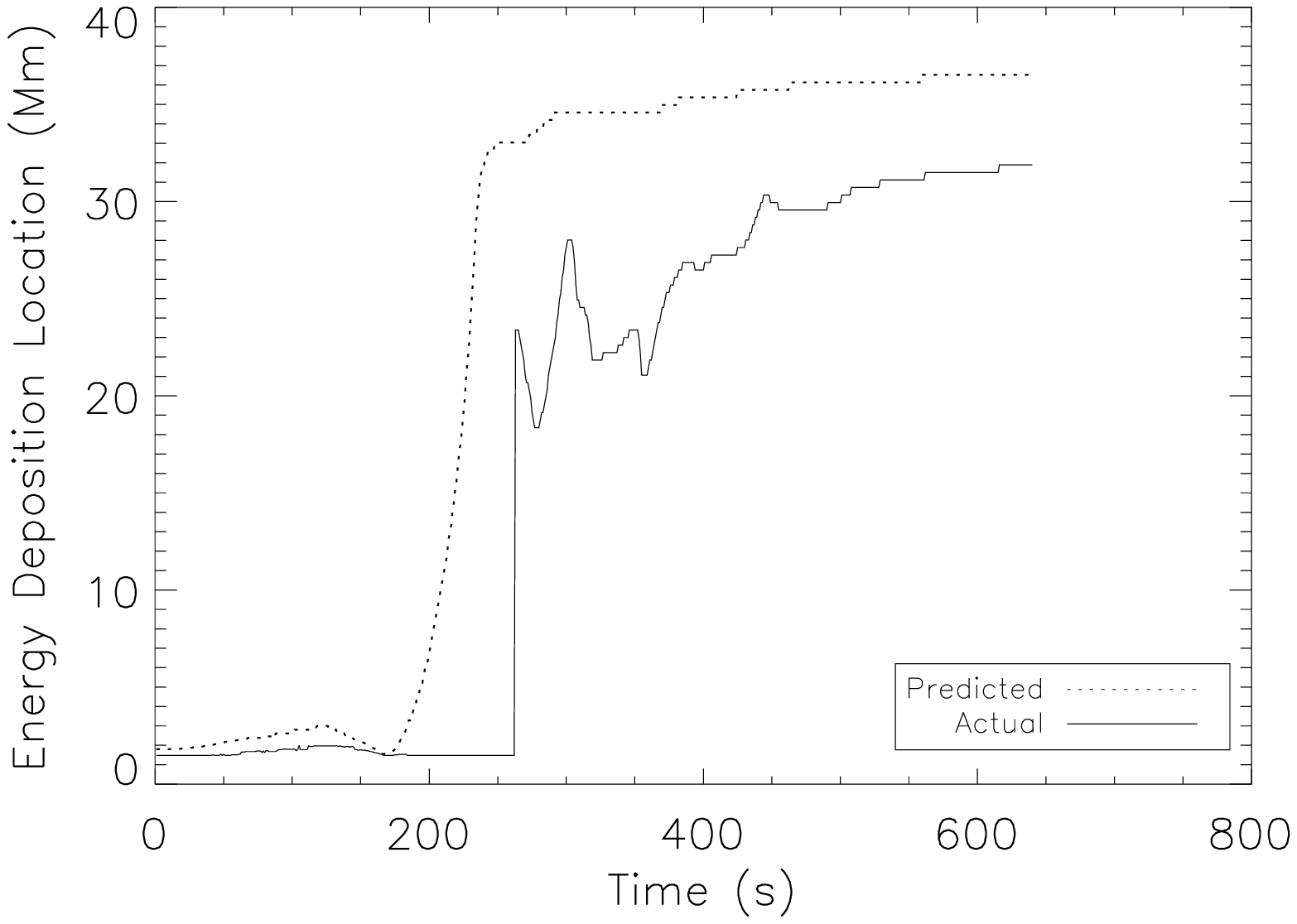}
\end{minipage}
\hspace{0.1in}
\begin{minipage}[b]{0.5\linewidth}
\centering
\includegraphics[width=3.25in]{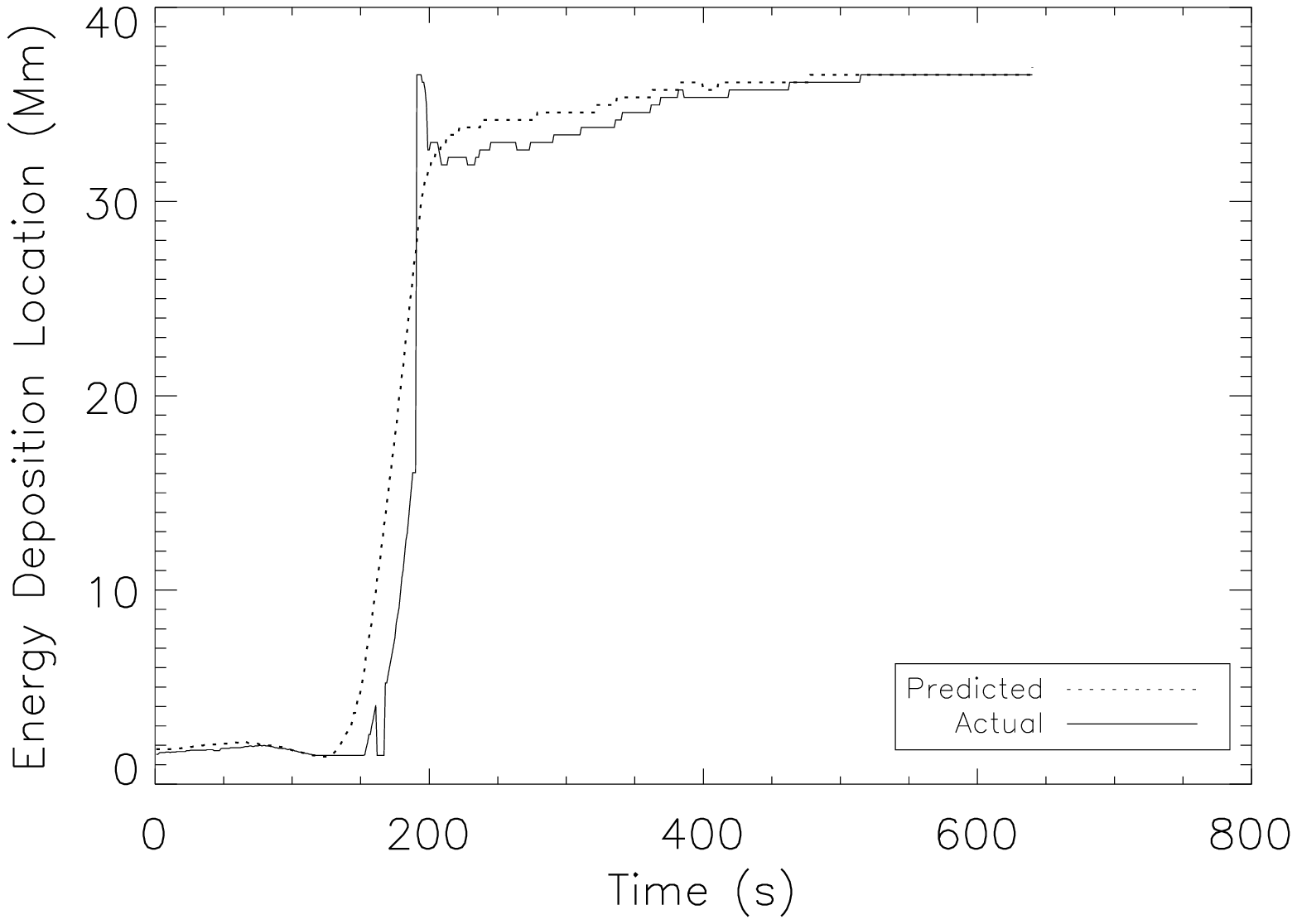}
\end{minipage}
\caption{The location of maximal energy deposition for Runs 1 and 20 (left and right, respectively) in one half of the loop.  The predicted curve uses the approximation for the mean stopping column density for an electron of energy $E_{c}$ \citep{nagai1984}.}
\label{deploc}
\end{figure}

The cut-off energy of the electron distribution can alter the GOES class significantly (see the third plot in Figure \ref{goesbeam}).  Slight decreases in the cut-off increase the GOES flux, but large decreases will significantly decrease the GOES flux.  On the other side, as the cut-off increases more and more, the energy is deposited deeper in the dense lower atmosphere, where there is a greater heat capacity and the energy is radiated away more quickly, leading to less heating, less thermal bremsstrahlung, and thus a lower GOES class in the 1-8 \AA\ channel.  In the 0.5-4 \AA\ channel, increases in the cut-off increase the amount of non-thermal bremsstrahlung, so the flux in this channel tends to increase with higher cut-off energies.  At extremely high cut-offs, the energy is deposited too low to heat the loop, so no thermal bremsstrahlung will be emitted; however, there will still be non-thermal emissions as the beam traverses the loop, which will be determined primarily by the energy flux in the beam.  Eventually, as the cut-off is increased to large enough values, the emission in both channels will be entirely non-thermal bremsstrahlung, to which the higher energy channel is more sensitive.  As the cut-off decreases, more heat is deposited higher up, leading to a higher temperature and more thermal emission.  If it decreases too much though, the beam is then essentially composed of thermal electrons, leading to less non-thermal bremsstrahlung emission and a lower GOES class (in both channels).  The cut-off energy is directly related to the location of energy deposition of the beam electrons.  \citet{nagai1984} give the mean stopping column density of an electron with energy $E_{c}$ as $\approx 10^{17} [E_{c}$ (keV)$]^{2}$ cm$^{-2}$, which can then be used to find the location of maximal energy deposition.  In Figure \ref{deploc}, we show the predicted and actual location of maximal energy deposition as a function of time for Runs 1 and 20 in one half of the loop.  In Run 20, it should be noted, the coronal density reached higher values and so the deposition location is in general higher than in Run 1, leading to more heating and more thermal bremsstrahlung, and thus a higher GOES class.  

\begin{figure}
\centering
\includegraphics[width=3.25in]{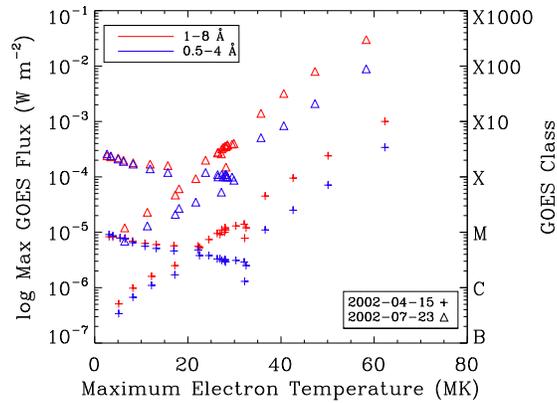}
\caption{The maximum electron temperature attained in the 60 simulations versus the maximum GOES flux, in each channel.  For both flares, there is a clear connection between the temperature and GOES flux in each channel, although the values differ between the two flares because of different loop lengths and cut-off energies.  The horizontal branches at low temperatures and relatively high GOES fluxes are the simulations with extremely high cut-off energies, where the energy is deposited too low in the atmosphere to heat the loops. }
\label{classtemper}
\end{figure}

There is also a clear relation between the maximum temperature of a flare and the maximum GOES flux.  Figure \ref{classtemper} shows the maximum electron temperature versus the maximum GOES flux found in the 60 simulations.  There is a clear correlation between the two, although there are two separate trends for each group of 30 simulations.  This is due to differences in the spectral index.  The index was much higher at all times in the M1.2 flare, and thus more energy was deposited higher in the atmosphere, leading to consistently higher temperatures despite lower GOES classes.  The spectral index therefore determines whether a flare will be thermally driven or beam dominated.  Note, however, that in each case there is a horizontal branch extending to lower temperatures with roughly constant GOES flux.  These branches are the cases where there are extremely high cut-off energies, which deposit their energy too low to heat the loop, but still produce roughly constant non-thermal emissions.

These results should be compared to Figure 6 and Equation 6 of \citet{feldman1996}, who found a similar trend, but significantly lower temperatures.  These lower temperatures result from the fact that they were measuring the temperature at the time of peak emission of the Fe XXV or Ca XIX channels of the Bent Crystal Spectrometer on SMM, which occurs after the time of maximum temperature (the density continues to rise).  For example, in Run 1, the maximum temperature was 260 seconds into the flare, while the maximum intensity of the Ca XIX line occurred 500 seconds into the flare, a 4-minute discrepancy, by which time the loop had cooled by 7 MK (see \citealt{bradshaw2011} for explanation of the forward modeling of spectral lines such as the Ca XIX line).  

There is some observational evidence for cold flares, which have detectable GOES emission but low temperatures.  For example, \cite{fleishman2011} report a C-class flare with a maximum temperature of 6 MK.  The authors suggest that the low temperature is due to a weak beam flux, which leads to little heating, so that the GOES emission would then be composed primarily of non-thermal bremsstrahlung.  This explanation agrees with our results, where small beam fluxes produce low temperatures, but emission detectable by GOES (see the lower left of Figure \ref{classtemper}).  One other possibility, which seems less likely, is that the flare had a large cut-off energy so that the energy was deposited deep in the chromosphere, once again leading to minimal heating but some emission in the GOES passbands nonetheless (see the horizontal branches in Figure \ref{classtemper}).

\section{Summary and Conclusions}
\label{conclusions}

We have investigated the effect of individual beam heating parameters on the GOES classification of solar flares.  Using two independent sets of observationally determined parameters, 60 numerical experiments have been performed, from which GOES light curves have been synthesized.  We have found clear trends between the beam parameters and the GOES flux.  First, the GOES classification strongly depends on the total beam energy ($\Psi \propto E^{\alpha}$ with $\alpha$ around 1.7 in the 1-8 \AA\ channel and around 1.6 in the 0.5-4 \AA\ channel).  This parameter dominates: the SXR spectrum is primarily determined by the amount of heat deposited in the solar atmosphere.  Second, the spectral index of the beam electron distribution does not significantly alter the GOES class (whether the cut-off is sharp or a knee).  There is a small positive correlation, though: an increase in $\delta$ slightly increases the GOES class.  Finally, the cut-off energy, which determines the mean location of energy deposition of the beam, affects the amount of both the thermal and non-thermal bremsstrahlung produced.  Thus, changes to the cut-off can either increase or decrease the GOES flux, but it is not a simple relation, unlike the other two parameters.  

As noted earlier, the Neupert effect predicts a linear relation between non-thermal energy and SXR flux ($\Psi \propto E$), which is in disagreement with the present results and those of \cite{warren2004}.  In this paper we have shown that the GOES flux strongly depends on the cut-off energy of the electron beam as well.  A change in the cut-off energy can alter the relation between non-thermal energy and SXR flux.  Consider the case of an extremely high cut-off energy where there is little heating and thus little thermal bremsstrahlung: all the emission will be non-thermal, and thus a linear relation would follow (Equation \ref{thicktargetbrem}).  Although it seems unlikely that such extremely high cut-off energies occur, it is clear that the cut-off energy can affect the relation.  

We can constrain the parameters as well.  For example, Runs 38 and 39 have a GOES class larger than any flare ever observed, and thus can be discarded as unlikely to occur, setting an effective upper limit on the non-thermal energy budget in flares (the largest ever seen with GOES was around X40, \citealt{brodrick2005}).  The total non-thermal energy in the 2002-07-23 flare was $2.6 \times 10^{31}$ erg \citep{holman2003}, so there is an effective upper limit around $\approx 7.5 \times 10^{31}$ erg (for similar cut-off energies).  The extremely high cut-off energies used in some of the runs were also unrealistic.  For example, Runs 55-60 are X-class flares with maximum temperatures beneath 10 MK, and densities too low to be observed, suggesting that such high cut-offs are not realistic (at least for the entire duration of the flare, there is some evidence that they may reach high cut-offs temporarily: \citealt{warmuth2009}).  The SXR spectrum is only one facet of emissions produced by solar flares, though.  

In future work, we will predict extreme ultraviolet (EUV) spectra to further constrain the model.  The properties of EUV lines, such as widths, Doppler shifts, and intensities, can be predicted from the model and directly compared with observations to help pin down signatures of the fundamental driving mechanism of solar flares.  For example, \citet{tsurutani2005} point out that while the 4 November 2003 flare had the largest GOES flux ever measured, the 28 October 2003 flare was more intense in the EUV.  The model developed here needs to be used to explain these differences as well.

One of the authors (JWR) was supported by NASA Headquarters under the NASA Earth and Space Science Fellowship Program - Grant NNX11AQ54H.  We thank the anonymous referee for detailed comments that substantially improved the paper.  We also thank our colleagues at the SHINE and SPD conferences this summer for their helpful discussions and suggestions.  

\appendix
\section{Coulomb logarithms of beam electrons}
\label{coulomb}
In the derivation of a heating function due to a beam of charged particles interacting with a hydrogen target, \citet{emslie1978} defines three effective Coulomb logarithms (the following analysis is based on \citealt{ricchiazzi1982}).  In the special case that the beam particles are taken to be electrons, the first (due to interactions of beam electrons and protons) is given by (Equation 8 of \citealt{emslie1978}):

\begin{equation}
\Lambda = \ln{ \frac{m_{0} v^{3}}{\nu e^{2}}} \approx \ln{ \frac{m_{e} v^{3}}{\nu e^{2}}} \approx \ln{ \frac{\pi^{1/2} m_{e}^{3/2} v^{3}}{n_{e}^{1/2} e^{3}}}
\end{equation}

\noindent where $m_{e}$ is the electron mass, $m_{0} = \frac{m_{e} m_{i}}{m_{e} + m_{i}} \approx m_{e}$ is the reduced mass of the system, $v$ is the electron velocity, $e$ is the electron charge, and $\nu$ is the plasma frequency ($ = (\frac{n_{e} e^{2}}{\pi m_{e}})^{1/2}$ for an electron number density $n_{e}$).  Taking the average electron energy to be $\langle E \rangle = \frac{1}{2} m_{e}v^{2}$, we can rewrite this as:

\begin{eqnarray}
\Lambda &=& \ln{ \Bigg[ 2 \langle E \rangle  \Big(\frac{2 \langle E \rangle}{m_{e}}\Big)^{1/2} \Big(\frac{\pi m_{e}}{n_{e} e^{6}}\Big)^{1/2}  \Bigg]} \nonumber \\
	    &=& \ln{ \frac{ 2^{3/2} \pi^{1/2}}{e^{3}}} + 1.5 \ln{\langle E \rangle} - 0.5 \ln{n_{e}} \nonumber \\
	   &\approx& 66.0 + 1.5 \ln{\langle E \rangle} - 0.5 \ln{n_{e}} 
\end{eqnarray}

\noindent Note that $\langle E \rangle$ must be in units of erg here.

The second logarithm, due to interactions between beam electrons and neutral hydrogen, is given by (Equation 12 of \citealt{emslie1978}):

\begin{eqnarray}
\Lambda^{\prime} &=& \ln{ \frac{m_{e} v^{2}}{1.105 \chi_{H}}} \nonumber \\
	    &=& \ln{ \frac{2 \langle E \rangle}{1.105 \chi_{H}}} \nonumber \\
	   &=& \ln{ \frac{2}{1.105 \chi_{H}}} + \ln{\langle E \rangle} \nonumber \\
	   &\approx& 25.1 + \ln{\langle E \rangle}
\end{eqnarray}

\noindent where $\chi_{H} = 13.6$ eV is the ionization energy of hydrogen.  

Finally, the third effective Coulomb logarithm is (Equation 20 of \citealt{emslie1978})

\begin{eqnarray}
\Lambda^{\prime \prime} &=& \ln{ \frac{v}{c \alpha}  } \nonumber \\
	&=& \ln{[ (\frac{2 \langle E \rangle}{m_{e}})^{1/2} \frac{1}{c \alpha}   ]} \nonumber \\
	&=& \ln{ \frac{2^{1/2}}{m_{e}^{1/2} c \alpha} } + 0.5 \ln{\langle E \rangle} \nonumber \\
	&\approx& 12.3 + 0.5 \ln{\langle E \rangle}
\end{eqnarray}

\noindent where $\alpha$ is the fine-structure constant.  The mean value of the energy can be found from the properties of the beam injection:

\begin{eqnarray}
\langle E \rangle &=& \frac{\int_{0}^{\infty} E \mathfrak{F}(E) dE}{\int_{0}^{\infty} \mathfrak{F}(E) dE} = \frac{\int_{0}^{E_{c}} K \frac{E^{3}}{E_{c}^{2}} dE + \int_{E_{c}}^{\infty} K \frac{E^{1 - \delta}}{E_{c}^{-\delta}} dE}{\int_{0}^{E_{c}} K \frac{E^{2}}{E_{c}^{2}} dE + \int_{E_{c}}^{\infty} K \frac{E^{- \delta}}{E_{c}^{-\delta}} dE} \nonumber \\
&=& \frac{ \frac{E_{c}^{2}}{4} - \frac{E_{c}^{2}}{2 - \delta}}{\frac{E_{c}}{3} - \frac{E_{c}}{1 - \delta}} = \frac{3}{4}\frac{\delta - 1 }{\delta - 2} E_{c} 
\end{eqnarray}

\noindent with the condition that $\delta > 2$.  

\section{Reversal of the Thick-Target Bremsstrahlung Double Integral}
\label{reverse}

Thick-target bremsstrahlung is evaluated with a double integral, Equation \ref{thicktargetbrem}, where the inner integral $\nu(\epsilon, E_{0})$ gives the photon yield for a single electron of initial energy $E_{0}$, current energy $E$, and losing energy at a rate $\frac{dE}{dt}$ through collisions with the ambient medium.  The entire electron distribution is then integrated to give the photon yield for all of the electrons in the beam, as a function of time and position.  However, the form of Equation \ref{thicktargetbrem}, and in particular the form of the cross-section $Q(\epsilon, E)$, does not allow for simple integration if a fully relativistic cross-section is used.  Many authors therefore use a simpler cross-section, such as the Kramer's or non-relativistic Bethe-Heitler cross-section, so that the inner integral becomes analytic, although precision will be lost in calculating bremsstrahlung emissions.  

However, reversing the bremsstrahlung integral has the advantage that the form of the cross-section does not matter.  When reversed, the new inner integral will be analytic for almost all forms of electron distributions that are assumed in thick-target models.  To this end, it is rather advantageous to reverse the integral into the form given in Equation \ref{reversebrem} (see also \citealt{kontar2011}).  We now give a full explanation of the evaluation of this integral, for the case of our assumed electron distribution (other forms will be similar in their evaluation). 

First, we note that the integral must be split into two parts depending on what the photon energy $\epsilon$ is relative to the cut-off energy $E_{c}$ (note that if the electron distribution has a high energy break, it must be split into three parts).  We begin with the case where $\epsilon \geq E_{c}$.   Disregarding preceding constants,

\begin{eqnarray}
I &\propto& \int_{E=\epsilon}^{\infty} E\ Q(\epsilon,E)\ dE \int_{E_{0} = E}^{\infty} E_{0}^{- \delta}\ dE_{0} \\ \nonumber
	&\propto& \int_{E=\epsilon}^{\infty} E\ Q(\epsilon,E)\ dE\ \Big[ E^{1-\delta}  \Big] \\ \nonumber
	&\propto&  \int_{E=\epsilon}^{\infty} E^{2-\delta}\ Q(\epsilon,E)\ dE 
\end{eqnarray}

\noindent One may then make a choice of cross-section, depending on what physical aspects are under consideration (see \citealt{koch1959} for an extensive list of possible choices).  For all but the simplest choices, this integral requires numerical evaluation, for which Gauss-Laguerre quadrature is highly efficient.

Next, we consider the case that $\epsilon < E_{c}$.  Then, the integral must be split into two parts, such that $\epsilon \leq E \leq E_{c}$ and $E_{c} \leq E < \infty$, and then summed together (since electrons of any energy greater than the photon energy contribute to the integral).  In the lower range,

\begin{eqnarray}
	I_{\mbox{low}} &\propto&  \int_{E=\epsilon}^{E_{c}} E\ Q(\epsilon,E)\ dE\ \Bigg[\int_{E_{0}=E}^{E_{c}} \frac{E_{0}
^{2}}{E_{c}^{2}}\ dE_{0}\ + \int_{E_{0}=E_{c}}^{\infty} \frac{E_{0}^{- \delta}}{E_{c}^{- \delta}}\ dE_{0}  \Bigg] \\ \nonumber
	&\propto& \int_{E=\epsilon}^{E_{c}} E\ Q(\epsilon,E)\ dE\ \Bigg[\frac{E_{c}}{3} - \frac{E^{3}}{3 E_{c}^{2}} + \frac{E_{c}}{\delta - 1} \Bigg] \\ \nonumber
	&\propto& \int_{E=\epsilon}^{E_{c}} E\ \Bigg( E_{c}\ \Big[\frac{\delta+2}{3(\delta - 1)}\Big] - \frac{E^{3}}{3 E_{c}^{2}}   \Bigg)\ Q(\epsilon,E)\  dE  
\end{eqnarray}

\noindent which can be evaluated efficiently with Gauss-Legendre quadrature, for any choice of cross-section.  In the upper range,

\begin{eqnarray}
I_{\mbox{high}} &\propto&  \int_{E=E_{c}}^{\infty} E\ Q(\epsilon,E)\ dE\ \int_{E_{0}=E}^{\infty} \frac{E_{0}^{- \delta}}{E_{c}^{- \delta}}\ dE_{0} \\ \nonumber
	&\propto& \int_{E=E_{c}}^{\infty} E^{2 - \delta}\ Q(\epsilon,E)\ dE
\end{eqnarray}

\noindent which is the same as the original case, except that the lower limit here is $E_{c}$.  As noted before, the latter two integrals must be summed in the case $\epsilon < E_{c}$.

\end{document}